\documentclass[aps,twocolumn,groupedaddress]{revtex4}
\usepackage{graphicx}
\usepackage{amsmath}
\usepackage{sidecap}

\newcommand{\beq}{\begin{eqnarray}}
\newcommand{\eeq}{\end{eqnarray}}

\newcommand{\dd}[2]{\fd{\partial #1}{\partial #2}}

\newcommand{\vp}{v_\perp}
\newcommand{\vv}{{\bf v}}

\newcommand{\Phit}{\Phi_{\parallel}}

\newcommand{\scr}[1]{_{\mbox{\protect\scriptsize #1}} }
\newcommand{\E}{{\bf E}}
\newcommand{\B}{{\bf B}}

\def\hblabel#1{\label{#1}}
\newcommand{\x}{{\bf x}}
\newcommand{\vz}{v_{\parallel}}
\newcommand{\fd}[2]{\frac{\displaystyle #1}{\displaystyle #2}}
\newcommand{\cE}{{\cal E}}

\newcommand{\pz}{{p_\parallel}}
\newcommand{\pp}{{p_\perp}}
\newcommand{\ud}{{\bf u}_{\delta}}
\newcommand{\udd}{{\bf u}_{D\delta}}
\def\eq#1{(\ref{eq:#1})}
\def\ddt#1#2{\partial #1 /  \partial #2}

\renewcommand\vec[1]{\mathbf{#1}}

\newcommand\J{\vec{J}}

\begin{document}


\title{Double layer electric fields aiding  the production of energetic flat-top distributions and superthermal electrons within the exhausts from magnetic reconnection.}


\author{J.~Egedal}
\affiliation{Department of Physics, University of Wisconsin-Madison, Madison, Wisconsin 53706, USA}

\author{W.~Daughton}
\affiliation{Los Alamos National Laboratory, Los Alamos, New Mexico
87545, USA}

\author{A.~Le}
\affiliation{Los Alamos National Laboratory, Los Alamos, New Mexico
87545, USA}

\author{A.~L.~Borg}
\affiliation{Norwegian Meteorological Institute, Blindern, 0313
Oslo, Norway}


\date{\today}

\begin{abstract}

Using a kinetic simulation of magnetic reconnection it was recently shown that magnetic-field-aligned electric fields ($E_{\parallel}$) can be present over large spatial scales in  reconnection exhausts. The largest values of
$E_{\parallel}$ are observed within  double layers. The existence of double layers in the Earth's magnetoshere is well documented. In our simulation their formation is triggered by large parallel streaming of electrons into the reconnection region. These parallel electron fluxes are required for maintaining quasi-neutrality of the reconnection region and increase with decreasing values of the normalized electron pressure upstream of the reconnection region, $\beta_{e\infty}=2\mu_0n_{e\infty}T_{e\infty}/B_{\infty}^2$. A threshold ($\beta_{e\infty}<0.02$) is derived for strong double layers to develop. We also document how  the electron confinement,  provided in part by the structure in $E_{\parallel}$, allows sustained energization by perpendicular electric fields ($E_{\perp}$). The energization is a consequence of the confined electrons'  chaotic orbital motion that  includes drifts  aligned with the reconnection electric field. The level of energization is proportional to the initial particle energy and therefore is enhanced by the initial energy boost of the acceleration potential, $e\Phit=e\int_x^{\infty} E_{\parallel} dl$, acquired by electrons entering the region. The mechanism is effective in an extended region of the reconnection exhaust allowing for the generation of superthermal electrons in  reconnection scenarios, including those with only a single x-line.  An expression for the phase-space distribution of the superthermal electrons is derived,  providing an accurate match to the kinetic simulation results. The numerical and analytical results agree with  detailed spacecraft observations  recorded during reconnection events in the Earth's magnetotail.
\end{abstract}


\maketitle

\section{Introduction}

It is widely believed that magnetic reconnection energizes electrons and is the source of superthermal electrons observed during solar flare events \cite{krucker:2010} and through {\sl in situ} measurements in the Earth's magnetosphere    \cite{oieroset:2001}.
Several previous studies of reconnection conclude that the dissipation  sites (where electrons are energized) are limited to electron scale regions and are insufficient in size to account for the large scale electron energization. To circumvent this conundrum recent models have invoked the possibility of multiple reconnection locations and flux-rope formation, which can facilitate energy release in a larger spatial volume and thereby help energize the large number of electrons required by observations \cite{drake:2006,oka:2010,hoshino:2012,drake:2013}.

While there exists clear evidence of magnetic flux-ropes forming in the Earth's magnetosphere, these flux-ropes do not appear to be volume filling such that energization at multiple x-lines becomes effective \cite{chen:2008,oieroset:2012}. This suggests that energization within the exhaust from a single x-line needs to be efficient to account for the extensive energization levels recorded experimentally. We therefore revisit electron energization in single x-line reconnection and find that superthermal electrons are produced over large spatial scales in the single x-line reconnection exhaust when the total pressure of the plasma upstream of the reconnection region is dominated by the magnetic field ($\beta_{e\infty}=2\mu_0p_{e\infty}/B_{\infty}^2\ll1$). In our kinetic simulation, a key to the strong electron energization is the break-down of adiabatic electron dynamics along the magnetic field lines, which, we find, is likely to occur when $\beta_{e\infty}<0.02$.

The documented heating mechanism and non-adiabatic parallel electron dynamics lead to the formation of strong magnetic field, aligned (parallel) electric fields, $E_{\parallel}$. These structures are consistent with double layers, where local charge separation generates strong parallel electric fields \cite{block:1978,singh:1987,raadu:1988}. The double layers
provide an initial energization of electrons streaming into the reconnection region (and exhaust) along field lines. In addition, the double layer $E_{\parallel}$ helps confine the electrons within the reconnection region, allowing for further energization by the reconnection electric field over the duration of multiple electron bounce orbits. The resulting electron distribution functions have characteristic signatures shaped by the interplay between the parallel and perpendicular electric fields during the energization process, which compare favorably to electron distributions observed {\sl in situ} by spacecraft in the Earth's magnetotail \cite{oieroset:2001,chen:2008jgr,egedal:2008jgr,egedal:2012}. As is the case for the magnetotail, $\beta_{e\infty}$ is also believed to be small upstream of reconnection sites associated with solar flares \cite{fletcher:2011}, where the electron confinement provided by $E_{\parallel}$  may also be important to explain the long lifetime of hard X-ray sources at flare loop-tops \cite{krucker:2010,li:2012}.

In previous work \cite{egedal:2008jgr,le:2009,egedal:2013} we have mostly analyzed scenarios including a guide magnetic field sufficient in strength that the magnetic moments, $\mu=m_e\vp^2/(2B)$, of the electrons are conserved, yielding  adiabatic (time reversible) electron dynamics for both the parallel and perpendicular electron dynamics.
Here we consider anti-parallel reconnection, which includes unmagnetized chaotic electron behavior in the reconnection exhaust leading to irreversible electron dynamics. More important for the energization process, however, is the aforementioned break-down of the adiabatic electron dynamics parallel to the magnetic field, causing the formation of large scale $E_{\parallel}$ regions. The spatial extent of these regions  is not tied to kinetic scales and expands with the reconnection exhaust. This suggests that the presented heating mechanism may be applicable to systems like solar flares, which are much larger than the simulation domain \cite{egedal:2012}.

The paper is organized as follows: In Sec.~II we discuss the formation of strong double layers,  which form along the reconnection separators. The double layers are responsible in part for the confinement of electrons in the exhaust and we derive a threshold for their development. In Sec~III we consider properties of the exhaust where pitch angle mixing is effective, and derive the rate for the energization of electrons trapped in the exhaust. In Sec.~IV we investigate the electron distributions that form within the exhaust, providing an explanation for the observed electron flat-top distributions. In addition, a model is derived for the superthermal electrons which is in good agreement with the kinetic simulations results. In Sec.~V the numerical and theoretical results are compared to spacecraft observations and the paper is concluded in Sec.~VI.

\section{Electron holes, and double layers}

\subsection{Electron pressure anisotropy, the acceleration potential, $\Phit$, and a condition for adiabatic electrons}

In this paper we document and analyze numerical simulation results obtained with the VPIC fully kinetic plasma simulation code \cite{bowers:2009}, implementing a configuration with $\beta_{e\infty}=0.003$. The  two-dimensional simulation is initialized with a Harris neutral sheet, and open boundary conditions are employed in the $x$ and $z$ directions. The initial particle distributions
include counter-drifting Maxwellian ion and electron populations localized to support the Harris current layer. In addition,  a separate background density is included, resulting in a total density profile $n(z)=n_0$sech$^2(z/d_i) +n_{\infty}$, where $n_0$ is the central Harris density, $n_{\infty}$ is a uniform background density and $n_{\infty}/n_0 =0.05$  for this simulation. The ion-to-electron temperature ratio of the Harris population is
$T_{i0}/T_{e0}=5$, whereas the temperatures, $T_{i\infty}$ and $T_{e\infty}$, for the uniform background populations have the same ratio, but are a factor of three colder for both species. Lengths are normalized by the ion $d_i=c/\omega_{pi}$ based on the Harris sheet density $n_0$, in a domain size of $L_x\times L_z =320d_i\times30d_i$. Other
parameters are $m_i/m_e =400$, $T_{i\infty}=m_ec^2/29$ and $T_{e\infty}=m_ec^2/144$, and $\omega_{pe}/\Omega_{ce}=2$, where $\Omega_{ce} = eB_{\infty}/(m_ec)$.  The simulation data presented in Ref.~\cite{egedal:2012} were also from this run, but here we provide a more in depth analysis and refined understanding of the numerical results.

\begin{figure}[h]\centering
  \includegraphics[width=9.0 cm]{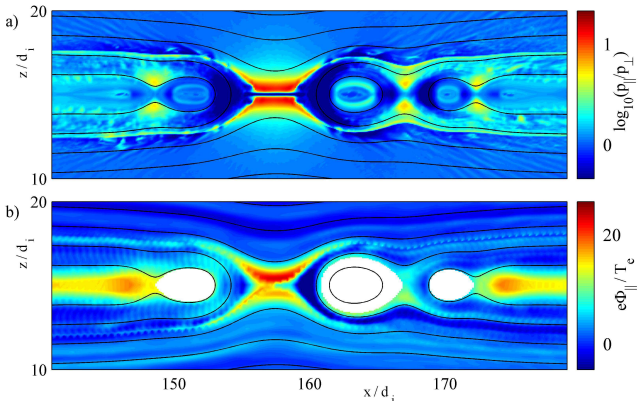}%
  \vspace{-0.5cm}
\caption{(a-c): Color contours of constant  pressure anisotropy $\log_{10}(\pz/\pp)$, and acceleration potential $e\Phit/T_{e\infty}$.
The magnetic field lines inside the island do not reach the ambient plasma and the acceleration potential, $\Phit$, is thus undefined within the areas left ``white'' in (b).
  }
\label{fig:early}
\end{figure}

Fig.~\ref{fig:early} shows the profiles of the electron pressure anisotropy $\pz/\pp$ and the acceleration potential for a relatively early stage of the reconnection process, $t\,\Omega_{ci}=31$.
Here $\Phit$   is defined as
\begin{equation}
 \hblabel{eq:Phit}
 \Phit=\int_x^{\infty}E_{\parallel} dl\quad,
\end{equation}
and is a pseudo potential which measures the integrated parallel electric field $E_{\parallel}$ along the magnetic field lines. Because of the large electron thermal speed the magnetic field lines can often be considered stationary during a single electron transit \cite{egedal:2013}. Thus, $e\Phit$ characterizes the energy that electrons acquire in their free streaming along magnetic field lines into the reconnection region \cite{egedal:2009pop}.

In Fig.~\ref{fig:early}, the relatively large pressure anisotropy and large values of $e\Phit/T_{e\infty}$ in the inflow are accurately described by the self-consistent adiabatic electron model derived in Refs.~\cite{egedal:2008jgr,le:2009,egedal:2013}. The model provides equations of state (EOS) for the electron pressure components $\pz=\pz(n,B)$ and $\pp=\pp(n,B)$. For anti-parallel reconnection the EOS are applicable to the inflow region where the electron magnetic moments, $\mu=m_e\vp^2/(2B)$, are conserved. The model includes the non-linear effects of electron trapping by the magnetic mirror force and by $\Phit$. For the present geometry with $e\Phit/T_{e\infty}\simeq18$, the trapped electrons dominate the properties of the electron fluid, and in this limit the electron pressure components $\pz\propto n^3/B^2$ and  $\pp\propto nB$ resemble the CGL scalings \cite{chew:1956}. In addition to the EOS, the model also yields the self-consistent values of $\Phit=\Phit(n,B)$.

Although the EOS do not apply to the electron diffusion region where the electrons become unmagnetized, through a momentum balance analysis of the electron layer it is shown that the EOS   regulate the integrated current across the layer \cite{le:2010grl,ng:2011}. Another result of the  momentum balance analysis in Ref.~\cite{le:2010grl} is a scaling law for the magnitude of $\Phit$ expected just upstream of the electron diffusion region:
\begin{equation}
\hblabel{eq:ephi} \left(\frac{e\Phit}{T_{e\infty}}\right)
\approx
\frac{1}{2}\left[\left(\frac{4\tilde{n}}{\beta_{e\infty}}\right)^{1/4}
- \frac{1}{2}\right]^2\quad,
\end{equation}
where the predicted values of $e\Phit/T_{e\infty}$ become large at small values of $\beta_{e\infty}$. Here $\tilde{n}=n/n_{\infty}$ and $\beta_{e\infty}=2\mu_0n_{\infty}T_{e\infty}/B_{\infty}^2$ is the normalized electron pressure upstream of the reconnecting current sheet.

Beside the break-down of the magnetic moment as an invariant, a second mechanism  also causes the adiabatic EOS to become invalid. This second mechanism is related to non-adiabatic effects in the parallel particle motion, and occurs at small values of $\beta_{e\infty}$. To estimate the critical value of $\beta_{e\infty}$ marking the transition to the non-adiabatic regime, we note that one important requirement for adiabatic parallel behavior is that changes in $\Phit$ for a flux-tube moving into the reconnection region must be small during an electron transit through the region. To quantify this condition we revisit the derivation of the adiabatic model in Ref.~\cite{egedal:2013}, where a key element for solving the drift kinetic equation is an assumed ordering
\[
\nabla_{\parallel}\sim \fd{1}{L}\,,\quad\nabla_{\perp}\sim\fd{1}{D}\,,\quad \dd{}{t}\sim\frac{v_D}{D},\]
\[\fd{D}{L}\sim\delta\,, \quad\frac{v_D}{v_{te\infty}}\sim\delta^2.
\]
Here $D$ and $L$ are the typical length scales across and along the reconnection region, respectively, with $D/L\simeq1/10$. The ratio between the electron drift speed $v_D\simeq0.1v_A$ and the thermal speed  is small $v_D/v_{te\infty}\simeq 1/400$ for $\beta_{e\infty}=1$, but increases with decreasing values of $\beta_{e\infty}$.

The adiabatic solution in Ref.~\cite{egedal:2013}  corresponds to  the limit where only electrons with small parallel
energy may become trapped. This requires that during an  electron transit time $\tau\simeq L/v_{te\infty}$ the changes in the magnetic and electric well (trapping electrons) must be small compared to $T_{e\infty}$. If this condition is not satisfied the parallel electron behavior may become non-adiabatic.
Following Eq.~13 of Ref.~\cite{egedal:2013}, the  condition for adiabatic dynamics is then
\begin{eqnarray}
\hblabel{eq:adia}
T_{e\infty} &>&\mu\dd{B}{t}\tau + e\dd{E_{\parallel}}{t}\tau L \nonumber \\[2ex]
&>&\mu B\fd{v_D}{D}\fd{L}{v_{te\infty}} + eE_{\parallel}\fd{v_D}{D} \fd{L}{v_{te\infty}} L\nonumber \\[2ex]
&>& \fd{v_A}{v_{te\infty}}\left( T_{e\infty} +e\Phit \right)\quad,
\end{eqnarray}
where we have used $v_A\simeq v_D L /D$.
For the low values of $\beta_{e\infty}$  and assuming $\tilde{n}=1$,  Eq.~\eq{ephi} is approximately $e\Phit/T_{e\infty}=1/\sqrt{\beta_{e\infty}}\,\,(\gg1) $.  Furthermore,  because $v_A^2/v_{te\infty}^2= m_e/(m_i\beta_{e\infty})$,
 Eq.~\eq{adia} may be written as $\beta_{e\infty}> \sqrt{m_e/m_i}$, as required to ensure parallel adiabatic behavior. Thus, we expect non-adiabatic parallel behavior in the inflow (and along the separators) for
 \begin{equation}
\hblabel{eq:adia3}
\beta_{e\infty} <  \sqrt{\fd{m_e}{m_i}}\quad.
\end{equation}
At the full mass ratio $m_i/m_e=1836$ this condition is $\beta_{e\infty}<0.02$, whereas for $m_i/m_e=400$ (applied in our simulations)  the derived threshold is $\beta_{e\infty}<0.05$.
 Again, the simulation studied here is  for a value  $\beta_{e\infty}=0.003$, and indeed, at later times in the
run strong non-adiabatic behavior is observed, resulting in values of $\pz$ and $e\Phit/T_{e\infty}$  much larger than those predicted by the adiabatic theory.

\subsection{Electron holes, double layers and the formation of a large amplitude acceleration potential}

\begin{SCfigure*}
  \includegraphics[width=14.0 cm]{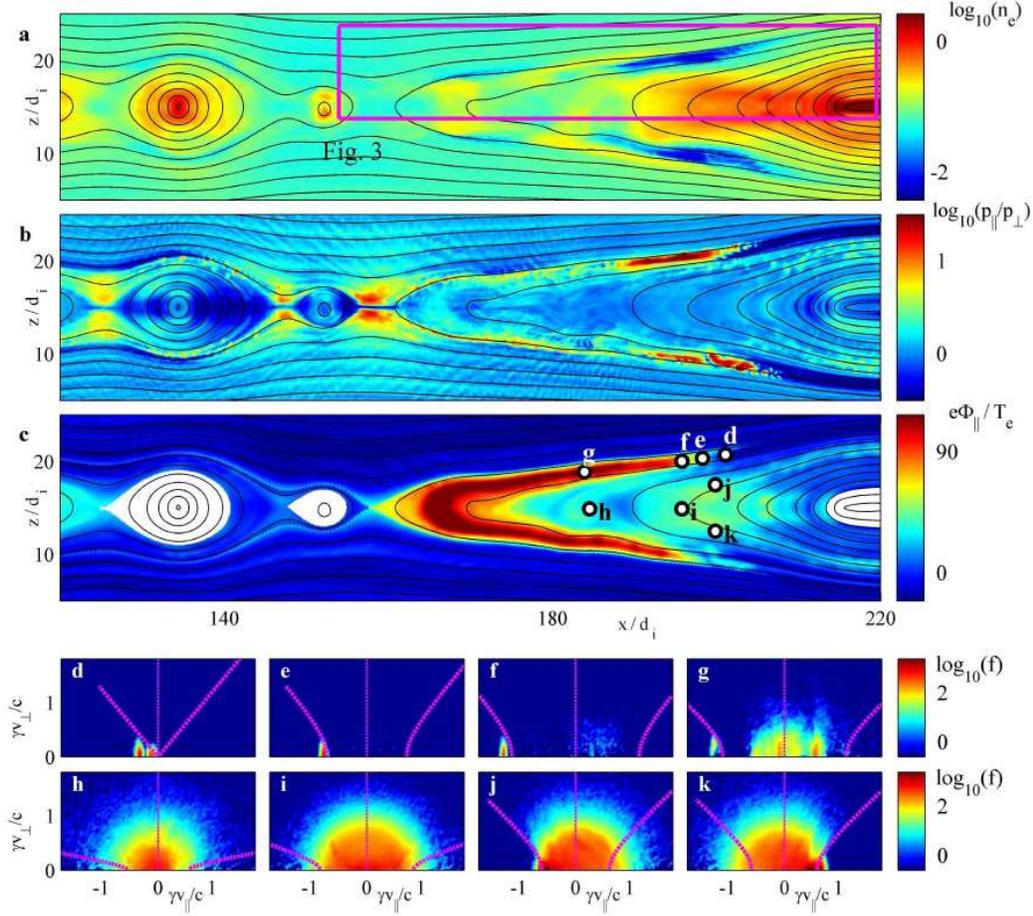}%
  \vspace{-0.5cm}
\caption{(a-c): Color contours of constant density $\log_{10}(n_e)$, pressure anisotropy $\log_{10}(\pz/\pp)$, and acceleration potential $e\Phit/T_e$.
Spatial locations are marked in (c) for which $f(\vz,\vp)$ is shown in panels (d) to (k). The magenta lines in (d) to (k) mark the trapped-passing boundaries as defined by Eq.~\eq{TPboundary}.
  }
\label{fig:G3}
\end{SCfigure*}

As mentioned above, the acceleration potential with  $e\Phit/T_{e\infty}\simeq 18$ in Fig.~\ref{fig:early}(b) recorded for $t\Omega_{ci}=31$ is consistent with the scaling law for $\Phit$ in Eq.~\eq{ephi}. In contrast, the acceleration potential for $t\Omega_{ci}=56$ displayed in Fig.~\ref{fig:G3}(c) has a much larger magnitude, $e\Phit/T_{e\infty}\simeq100$, and is caused by the formation of structures  in $E_{\parallel}$ for which the largest amplitudes are observed within density cavities.  The associated jumps in $\Phit$  are consistent with double layers \cite{block:1978,singh:1987,raadu:1988}.
The existence of such electron holes and double layers  is well documented by spacecraft observations within the magnetosphere. In the early observations they were termed {\sl broadband electrostatic noise} \cite{gurnett:1997} but Geotail observations showed they are
solitons and not monochromatic mixture or coherent broadband tones \cite{matsumoto:1994}. More modern and detailed observations include events recorded by the THEMIS mission \cite{ergun:2009, andersson:2009} and the Van Allen probes \cite{mozer:2013}.

 Two density  cavities (with strong double layers) are clearly seen near the separators in Fig.~\ref{fig:G3}(a) for $x/d_i\simeq200$.
In Fig.~\ref{fig:G3}(c) four points, $d$ -- $g$, are selected on a field line sampling the center of a large amplitude structure in $\Phit$. The corresponding distributions are shown in  Figs.~\ref{fig:G3}(d--g). For point $d$ the value of $e\Phit/T_{e\infty}\simeq 5$ is relatively small;  the distribution mainly consists of a beam of incoming electrons. Likewise for points $e$ and $f$, the distributions are almost purely composed of the incoming beams energized by $\Phit$ with velocities reaching $\gamma \vz/c\simeq -1.2$. For point $g$, in addition to the incoming beam, a population of  electrons are observed mostly with velocities $\gamma v/c < 0.7$. These relatively hot electrons originate from the reconnection region, but given the large amplitude of $\Phit$, they are deeply trapped and cannot escape the reconnection region by parallel streaming along field lines.

\begin{SCfigure*}
  \includegraphics[width=12 cm]{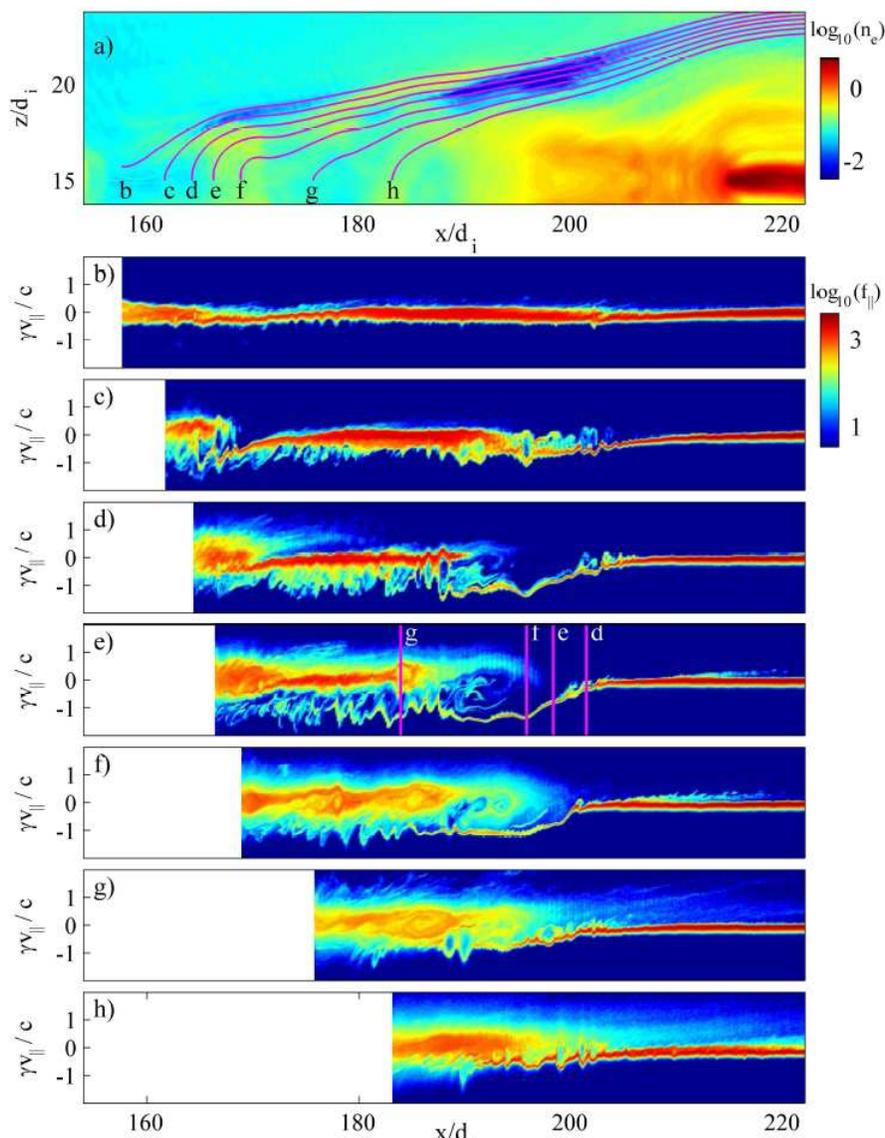}%
  \vspace{-1cm}
\caption{a) Color contours of constant density, $\log_{10}(n_e)$. Strong density cavities are observed where the density is reduce by a factor of 10 below the upstream value. Seven field lines are selected and marked $b$ to $h$. Along each field line the parallel distribution is computed, $f_{\parallel}(\vz)=2\pi \int f \vp d\vp$, and displayed in panels b) to h) as functions of $\gamma\vz$ and $x/d_i$. The magenta lines in e) marked $d$--$g$ indicate the positions corresponding to the distributions in Fig.~\ref{fig:G3}(d--f), respectively.}
\label{fig:dispAling56}
\end{SCfigure*}

To provide more details on these large scale structures yielding direct acceleration of the incoming electrons, in Fig.~\ref{fig:dispAling56}(a) a zoom-in-view is given for the region outlined in Fig.~\ref{fig:G3}(a).
Within this region seven field lines are selected for which $f_{\parallel}(\x,\vz) =2\pi \int \vp f\,d \vp$ are computed. The resulting $f_{\parallel}(\x,\vz)$ are displayed in Figs.~\ref{fig:dispAling56}(b-h) as functions of $x$ along the selected field lines.
The upper most field line $(b)$ in Fig.~\ref{fig:dispAling56}(a) is just upstream of the density cavities. The corresponding $f_{\parallel}$ along this field line is shown in Fig.~\ref{fig:dispAling56}(b). Signatures of developing instabilities are seen for $x/d_i\simeq170$.

Strong instabilities  are visible  along the second field line $(c)$ for which $f_{\parallel}$   displays evidence of  both electron holes and double layers.
The electrons with $\gamma\vz>0$ are streaming away from the reconnection region. The first double layer for  $x/d_i\simeq170$ is seen to help confine electrons energized in the X-line region. However, the second double layer at $x/d_i\simeq200$ develops without the presence of energized electrons from the reconnection site.

Figs.~\ref{fig:dispAling56}(d,e) illustrate how the double layers continue to develop. The vertical magenta lines marked $d$ -- $g$ in Fig.~\ref{fig:dispAling56}(e) correspond to the distributions in Figs.~\ref{fig:G3}(d-g), respectively. The beams  in Figs.~\ref{fig:G3}(d-g) are readily identified as the structure (with $\gamma \vz < 0$) in  $f_{\parallel}$  containing incoming electrons  continuously accelerated by $\Phit$. For these field lines at the center of the density cavity, $e\Phit$ is larger than the energies of electrons streaming away from the reconnection site. Thus, for $x/d_i>200$ almost no escaping  electrons  (electrons with $\gamma \vz>0$) are observed.

The magnitude of the double layer in Fig.~\ref{fig:dispAling56}(f) is still large. In addition to the incoming beam, the double layer is now being filled with energetic electrons streaming away from the reconnection site and reflected back again by the double layer. Considering the evolution of $f_{\parallel}$ in Figs.~\ref{fig:dispAling56}(b-f) it is clear that the acceleration by $\Phit$ of the incoming electron beam and the subsequent mixing in electron holes is responsible for the main heating of the electrons on these field lines intersecting the density cavities.

Deeper into the exhaust the value of $\Phit$ is reduced. This is evident in Figs.~\ref{fig:dispAling56}(g,h) where the incoming electron beam  now has velocities on the order of $\gamma \vz/c\simeq -0.8$. The reduced values of $\Phit$ allow energized electrons to escape the reconnection region
and finite values of $f_{\parallel}$ are thus observed for $x/d_i>200$ with $\gamma \vz>0$. These escaping electrons carry a significant heat flux away from the reconnection region.

\subsection{Formation of strong double layers}

Recently,  Li {\sl et al}.~\cite{li:2012,li:2013,li:2014} investigated the formation of double layers at the interface between hot electrons energized at the reconnection site and  cold electrons streaming into the reconnection region along field lines. Their numerical studies  document electron holes and double layers  with signatures similar to those observed in our simulations outside the density cavities (i.e.~in Figs.~\ref{fig:dispAling56}(g,h)). They modelled the hot electrons from the reconnection region as a Maxwellian with a temperature $T_h$ and observed  double layer amplitudes, $e{\Phit}\scr{DL} \simeq 0.73 T_h$, less than $T_h$. This moderate amplitude  allows for a significant fraction of the energetic electrons to escape \cite{li:2012}.

Interestingly, the strong double layers that form within the density cavities in  Figs.~\ref{fig:dispAling56}(c-f) appear to have  some characteristics different from those studied by Li {\sl et al}..
For example, we here observe an amplitude much larger than the energy of electrons energized at the reconnection site, allowing no electrons to escape by parallel streaming along field lines. Furthermore, the development of these large amplitude double layers is not  driven by hot electrons streaming away from the reconnection region. This is clear because the strong double layer structure at $x/d_i\simeq200$ in  Figs.~\ref{fig:dispAling56}(c-d)) develops before energetic electrons from the reconnection region reach this location (about 40 $d_i$ downstream of the x-line).
Below we argue that the strong double layers (distinct  from the double layers outside the density cavities) form not just to reduce free streaming losses, but primarily to boost the electron density within the reconnection region.

\begin{figure}[h]\centering
  \includegraphics[width=9.0 cm]{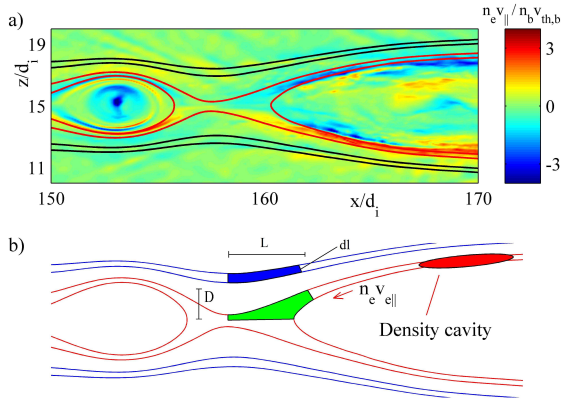}%
  \vspace{-0.5cm}
\caption{a) Contours of parallel electron flows into the reconnection region. b) Schematic illustration of a flux tube section (blue) expanding as it enters the reconnection region (green). To maintain a uniform electron density (required for quasi neutrality) electrons flow along field lines into the green region driving the formation of the density cavity marked in red.   }
\label{fig:jpar}
\end{figure}

From the principle of quasi-neutrality, electric fields in a plasma develop to maintain near equal densities of electrons and ions.  To elucidate the mechanisms driving the strong double layers, in Fig.~\ref{fig:jpar}(a) contours of the parallel electron flux are shown for the reconnection region, and large fluxes of electrons $|n_eu_{e\parallel}|/n_{\infty}v_{te\infty}\simeq4$ flow toward the reconnection region. Thus, this flow is stronger than the incoming parallel flux of upstream electrons $\Gamma_{\parallel\infty} = 2\pi \int_{0}^{\infty} d\vz \int_0^{\infty} \vp d\vp \vz f_{\infty} = n_{\infty}v_{te\infty}/(2\sqrt{\pi})$.

To understand why the  parallel flux of electrons is  needed, we  consider the blue area, $A_1 (= L\,dl)$ on the schematic flux-tube in Fig.~\ref{fig:jpar}(b). During the reconnection process this area convects into the larger green area $A_2$. From magnetic flux conservation within the flux-tube, the areas of the blue and green regions are related as
\[A_2 = A_1 \fd{B_1}{B_2}\quad,
\]
where $B_1$ and $B_2$ are representative magnetic field strengths within the two areas. As the in-plane magnetic field vanishes at the X-line, as is  ubiquitous in reconnection we have $B_1>B_2$. The areas considered are within the ion diffusion region where the ions are decoupled from the magnetic field lines. The simulation shows that within this region the ion density, $n$, is nearly uniform, such that area $A_2$ includes an increased number of ions compared to area $A_1$. This increase can be estimated as
\begin{equation}
\hblabel{eq:Deln}
\Delta N = n (A_2-A_1) = n L \,dl \left(\fd{B_1}{B_2} -1\right)\quad.
\end{equation}
Meanwhile, the electrons are {\sl frozen in} to the magnetic field in their  perpendicular motion, so the matching increase $\Delta N$ of electrons (required for quasi neutrality) must be supplied by a parallel electron flux $n_e u_{e\parallel}$.

To estimate $u_{e\parallel}$ we consider again the characteristic length scales $D$ and $L$ across and along the reconnection region. Previous studies (including Ref.~\cite{shay:2004}) have established that the inflow velocity upstream of the ion diffusion region is $(D/L) v\scr{A}$, with $D/L\simeq0.1$. Meanwhile, inside the ion diffusion region the velocity of the flux-tubes are enhanced by the factor $B_1/B_2$
yielding $v\scr{in,e}=v\scr{A} B_1 D/(LB_2)$. Thus, region $A_1$ with length $L$ convects into region $A_2$ during a time $\tau\simeq D / v\scr{in,e} = LB_2/(v\scr{A}B_1)$.
The increase in the number of electrons within the considered area can then be
estimated as $\Delta N=n u_{e\parallel} dl\,\tau $. Equating this with Eq.~\eq{Deln}
 the factors of $n$, $L$ and $dl$ cancel such that
\[
u_{e\parallel} = v\scr{A}\fd{B_1}{B_2}\left(\fd{B_1}{B_2} -1\right) \simeq 4 v\scr{A}\quad,
\]
where we have assumed a ratio $B_1/B_2\simeq2.5$. This result may be rewritten on the form
\begin{equation}
\hblabel{eq:upar}
\fd{u_{e\parallel}}{v_{te\infty}}=4\sqrt{\fd{m_e}{m_i}}
\fd{1}{\sqrt{\beta_{e\infty}}}
\quad,
\end{equation}
so for the present simulation (and in agreement with Fig.~\ref{fig:jpar}(a)) we  then obtain $u_{e\parallel}\simeq 4 v_{te\infty}$.

In kinetic and also Hall MHD simulations of reconnection at larger values of $\beta_{e\infty}$ less pronounced  density cavities are typical along the separators, and it has been argued that they develop to maintain pressure balance perpendicular to separators  \cite{shay:2001}. We here emphasize the role of electrons streaming into the reconnection region
as the more important cause for the near depletion of electrons from the  density cavities.  To the best of our knowledge, this additional mechanism for density depletion   is described here for the first time.
We also note that the resulting pattern of the parallel electron currents along the separators are largely responsible for the characteristic Hall magnetic field of the reconnection region \cite{le:2014}.

 The development of the observed strong double layers is likely to be suppressed when
$n_{\infty}u_{e\parallel}<  \Gamma_{\parallel\infty} = n_{\infty}v_{te\infty}/(2\sqrt{\pi})$, allowing thermal streaming of electrons to supply the reconnection region with the electrons needed. With Eq.~\eq{upar}, a crude criteria for the development of the strong double layers is  then
\begin{equation}
\hblabel{eq:double}
\beta_{e\infty} \lesssim 200\, \fd{m_e}{m_i}
\quad.
\end{equation}

\begin{SCfigure*}
  \includegraphics[width=13.0 cm]{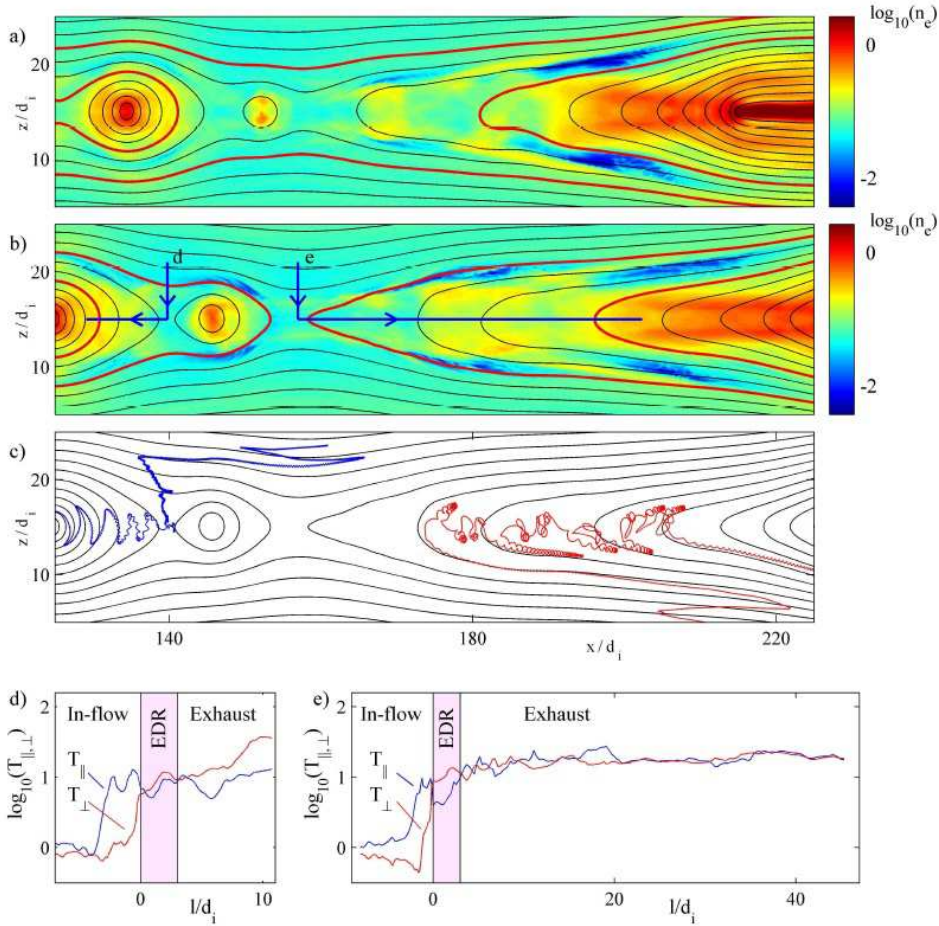}%
  \vspace{-0.5cm}
\caption{a,b): Contours of constant density for $t\Omega_{ci}=56$ and $t\Omega_{ci}=61$, respectively. c) Typical electron test orbits trajectories. In the left side exhaust the electron magnetic moments are conserved, whereas the chaotic electron motion in the right side exhaust is non-adiabatic. d,e): Electron temperature components evaluated along the lines shown in b).      }
\label{fig:pltOrbits}
\end{SCfigure*}

For $m_i/m_e=400$ this criteria suggests that the strong double layers may already form at $\beta_{e\infty}<0.5$. However, the derivation of Eq.~\eq{double} does not include the effects of the adiabatic $\Phit$ given in Eq.~\eq{Phit} helping to boost the electron flow into the reconnection region. In contrast, the condition in Eq.~\eq{adia3} includes the flows driven by $\Phit$ and therefore provides a more  accurate  threshold for the transition to the non-adiabatic parallel dynamics. Nevertheless, we include the above derivation of Eq.~\eq{double} as it elucidates  the mechanisms we believe are important for the formation of the density cavities and the associated strong double layer formation.

\section{Properties of the pitch angle mixed exhaust}
\subsection{Signatures of the pitch angle mixed exhaust}

The dynamics in the right side exhaust (for $x/d_i>160$) in Fig.~\ref{fig:G3} are different from the left side not only because of the asymmetry in $\Phit$. Another main difference is that within  the right side exhaust  the magnetic moments of the electrons are not conserved when the electrons cross the mid-plane. The distributions for the points marked  $h$ -- $k$ in Fig.~\ref{fig:G3}(c) are shown in Figs.~\ref{fig:G3}(h-k). The points $h$ and $i$ are both at the mid-plane and we notice how their distributions are independent of the pitch angle $\theta=\angle(\vv,\B)$. This is the signature of complete pitch angle mixing.

The magnetic moment, $\mu= m\vp^2/2$, is only an adiabatic invariant of the electron motion when the radius of curvature of the magnetic field $R_c$ is larger than the electron Larmor radius $\rho_l = m\vp/(eB)$. This requirement for adiabatic motion may also be expressed as $\kappa^2 = R_c/\rho_l > 1$ along the full electron trajectory \cite{buchner:1989}.
In the center of the exhaust the strongly bent field lines in combination with the low magnetic field strength, can lead to non-adiabatic electron motion with $\kappa^2<1$.
 This causes electrons to pitch angle mix,  washing  out anisotropic structures in velocity space.

To further explore the temporal evolution of the plasma, in Figs.~\ref{fig:pltOrbits}(a,b) density profiles for $t\Omega_{ci}=56$ and $t\Omega_{ci}=61$ are given, respectively. The main density cavities of $t\Omega_{ci}=56$ are seen to move downstream with the exhaust while new cavities form closer to the x-line. During this downstream propagation the cavities remain characterized by large values of $E_{\parallel}$, such that the region of parallel energization is expanding in time.

Examples of test particle orbits are given in Fig.~\ref{fig:pltOrbits}(c). For the part of the blue  trajectory which falls within the left side exhaust, the magnetic moment is an adiabatic invariant. Meanwhile, for the red trajectory typical of the right side exhaust, the magnetic moment is not conserved where the particles pass through the midplane of the domain.

The described  particle motion has  direct implications on the pressure profiles. This is evident in Figs.~\ref{fig:pltOrbits}(d,e) where the parallel and perpendicular pressure components are evaluated along the lines marked $d$ and $e$ in Fig.~\ref{fig:pltOrbits}(b). Consistent with the EOS by Le {\sl et al.}~for both inflow regions we observed $T_{e\parallel}\gg T_{e\perp}$. At the end of the electron diffusion regions pitch angle mixing leads to $T_{e\parallel}= T_{e\perp}$. For  the left side, the magnetic islands lead to an increase of the field strength in the exhaust such that the magnetic moments of the electrons are conserved, and  the betatron heating is effective. This  leads to the temperature components with $T_{e\perp}>  T_{e\parallel}$, and  is in contrast to the right side exhaust where pitch angle mixing yields $T_{e\perp} \simeq  T_{e\parallel}$.


\subsection{Simulation profiles of the pitch angled mixed exhaust}

\begin{figure}[h]
\centering
  \includegraphics[width=9.0 cm]{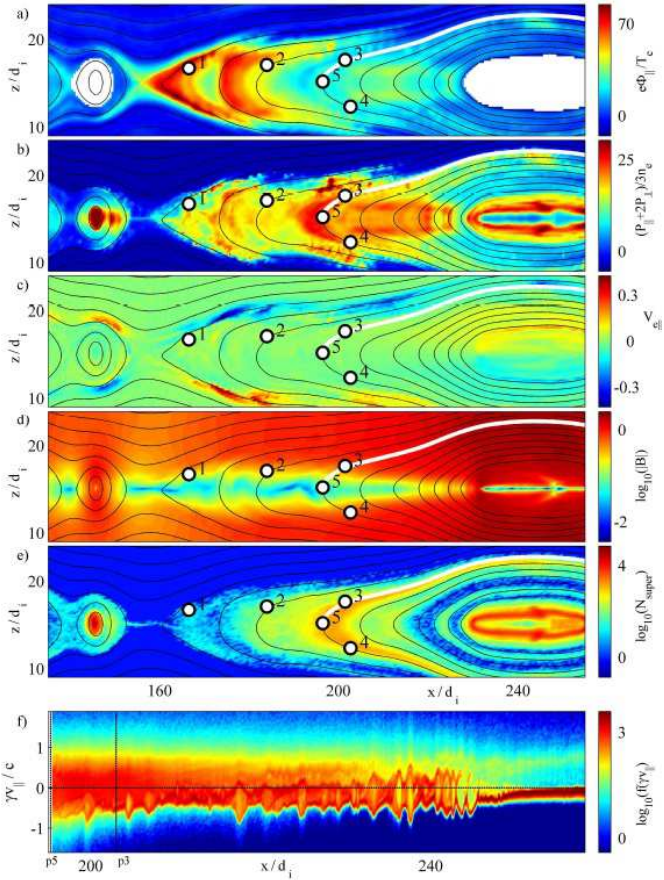}%
  \vspace{-0.5cm}
\caption{Kinetic simulation results showing contours of a) $e\Phit/T_e$, b) $T_e=Tr({\bf P}_e)/3n_e$, c)$V_{e\parallel}$ , d) $\log_{10}(B)$, and e) $n\scr{super}/n$ representing the density of electrons with $\gamma v/c>2.5$.  f) parallel phase-space distribution $f_{\parallel}(\gamma v_{\parallel}) $ as a function of $x$ along the field-line highlighted in white in a)-e). The simulation results shown correspond to time
$t\Omega_{ci}=61$.}
\label{fig:F1}
\end{figure}

Key profiles for the electron dynamics at $t\Omega_{ci}=61$ are given in Fig.~\ref{fig:F1}. The area of a significant acceleration potential ($e\Phit/T_{e\infty}>30$) in Fig.~\ref{fig:F1}(a) is much larger than at previous time, $t\Omega_{ci}=56$, considered in Fig.~\ref{fig:G3}(c). In the right side exhaust, the boundaries for large values of $e\Phit/T_{e\infty}\, (>30)$ coincide with the boundaries where the effective electron temperature, $T_e=(\pz+2\pp)/3n_e$, in Fig.~\ref{fig:F1}(b), displays an abrupt increase (by about a factor of 20) from its value in the inflow. This is consistent with our conclusions above that the $E_{\parallel}$ of the electron holes and  double layers provides strong energization of the incoming electrons.
This parallel heating is most significant within  the density cavities, which, as displayed in Fig.~\ref{fig:F1}(c), are characterized by the strongest streaming of electrons  directed towards  the reconnection region.

As discussed above, due to the low values of $B$ observed in Fig.~\ref{fig:F1}(d) along the exhaust mid-plane, the electron motion in the right side exhaust is non-adiabatic. Although the pitch angle mixing is limited to this region of low $B$, due to the non-localized motion of the trapped electrons, the pitch angle mixing impacts the electron distributions in the full width of the exhaust. We also note, that in contrast to the bulk electron heating, the generation of superthermal electrons occurs gradually in the exhaust. This is evident in Fig.~\ref{fig:F1}(e) where density contours are given for superthermal electrons with $\gamma v/c> 2.5$.

The parallel electron dynamics along a typical exhaust field line is illustrated in
Fig.~\ref{fig:F1}(f). As in Fig.~\ref{fig:dispAling56}, the contours represent constant values of $f_{\parallel}(v_{\parallel})= 2\pi\int f  \vp d\vp$, here as a function of $x$ along the field line highlighted in white in Figs.~\ref{fig:F1}(a-e). Again, large electron hole structures are observed, especially at the interface (for $x/d_i\simeq240$) between the cold incoming electrons and electrons already heated in the exhaust.

\subsection{Relative importance of $E_{\parallel}$ and $E_{\perp}$ for electron energization in the pitch angle mixed exhaust}

\begin{figure}[h]
\centering
  \includegraphics[width=6 cm]{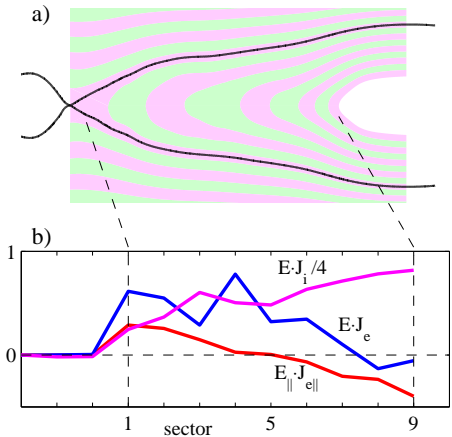}%
\caption{ a) The cross-section is sectored into intervals of $A_y$. For each sector $E_{\parallel} J_{e\parallel}$, $\E\cdot{\bf J}_e,$ and $\E\cdot{\bf J}_i,$ is evaluated as displayed in b).
}
\label{fig:F2}
\end{figure}

It is interesting to explore the relative importance of parallel and perpendicular electric fields for energizing the exhaust electrons.
The local energy gain of the electron fluid is quantified by $\E\cdot\J_e$. As shown in Fig.~\ref{fig:F2}(a) we divide a part of the simulation domain into sectors of equal increments in the in-plane magnetic flux, $A_y$. The energy exchange terms $E_{\parallel} J_{e\parallel}$, $\E\cdot\J_e$, and $\E\cdot\J_i$ are integrated over each sector and displayed in Fig.~\ref{fig:F2}(b).
Within the sector including the separator and the following two downstream sectors, $E_{\parallel} J_{e\parallel}$ amounts to about 50\% of the total $\E\cdot\J_e$.
As shown in Fig.~\ref{fig:F1}(e), these sectors include regions of strong electron flows $V_{e\parallel}$ toward the X-line. Because $\Phit$ is generally increasing as the X-line is approached, incoming electrons are  energized by $E_{\parallel}$.

Further downstream of the X-line the electron flow is away from the X-line (see Fig.~\ref{fig:F1}(c)), such that the electrons endure a net loss of energy through work against $E_{\parallel}$. In fact, the average value of $E_{\parallel} J_{e\parallel}$ in the exhaust is small, while the heating by $\E_{\perp}\cdot \J_{e\perp}$ accounts for nearly all the electron energization when integrated over the exhaust. Also noteworthy, as can be seen in the traces in Fig.~\ref{fig:F2}(b) the electrons acquire about 20\% of the total dissipated magnetic energy, which is large compared to levels typically observed in simulations at higher $\beta_{e\infty}$.

\subsection{Rate of electron energization by the reconnection electric field}
\label{heating}
It is natural to suspect the reconnection electric field, $E\scr{rec}$, to be the cause of the documented energization by $\E_{\perp}$. The rate of this energization is given by $P=E\scr{rec}J_{ey}$, where $J_{ey}$ is the electron current in the y-direction.
The strongly curved field lines of the exhaust are favorable for energization related to the curvature drift \cite{dahlin:2014}. This is shown directly in appendix A where we apply the guiding center model and obtain a Fermi-like energization rate proportional to the particle energy. However, it is not obvious that the guiding center model is applicable to the present exhaust where the magnetic moments of the electrons are not conserved. This motivates a more general derivation of the heating rate without the assumption of adiabatic electron motion. Therefore, we will estimate  the heating rate using a more general formulation. This estimate will also include the possible contributions from the magnetization currents (normally neglected in calculations  based on the guiding center model \cite{dahlin:2014}). The end result is the confirmation of the heating rates obtained with the guiding center model.

Our generally valid approach for estimating the heating rate of electrons in the pitch angle mixed exhaust is first to  generalize the MHD-force balance equation, ${\bf J_{\delta}}\times\B_{\delta}=\nabla p_{\delta}$, where the subscript $\delta$ denotes a particular class of electrons with $v\simeq v_0$ for $t=t_0$.
We may thereby determine $J_{\delta y}$ as a function of the electron energy, which in turn will provide us with the energization rate due the to the reconnection electric field as a function of energy.

To limit the analysis to electrons with an initial velocity around $v_0$ at time $t_0$, we introduce $f_{\delta}(\x,\vv,t)$ which at $t=t_0$ is given by
\begin{equation}
\hblabel{eq:fd}
f_{\delta}(\x,\vv,t) =
\left\{\begin{array}{ccc}
    f(\x,\vv,t) &, & v_0-\delta v < |\vv| < v_0+\delta v \\[2ex]
    0 &, & \mbox{elsewhere} \\
\end{array}\right.\quad,
\end{equation}
Naturally, the time evolution $f_{\delta}$ is governed by the Vlasov equation
\begin{equation}
\hblabel{eq:Vlasov}
\dd{f_{\delta}}{t} + \vv \cdot \nabla_{x} f_{\delta}  - \fd{e}{m_e}(\E + \vv\times\B)\cdot \nabla_{v} f_{\delta}=0
\quad,
\end{equation}
and moments over $f_{\delta}$ are defined in the regular fashion as
\[
\left<\vv^l\right>_{\delta} = \int \vv^l f_{\delta}\, d^3v \quad,
\]
yielding $n_{\delta}$, $n_{\delta}{\bf u}_{\delta}$ and ${\bf P}_{\delta}$ for $l=0, 1,$ and 2, respectively.

 In analogy with the derivation of the standard electron momentum equation, by taking the $l=1$-moment (or the ${\vv}$-moment) of Eq.~\eq{Vlasov} we obtain a momentum equation for the selected electrons
 \[
 m_e n_{\delta}\fd{d\ud}{dt}= -e n_{\delta}(\E+\ud\times\B)-\nabla p_{\delta} -\nabla\cdot {\bf \pi_{\delta}}\quad.
 \]
Here we have split the pressure tensor, ${\bf P}_{\delta}$, into its scalar part $p_{\delta}{\bf I}$ and the shear stress $\pi_{\delta}$. For the pitch angle mixed exhaust, the latter is assumed to be negligible, $\pi_{\delta}\simeq0$. For $t=t_0$, all electrons contributing to $p_{\delta}$ have velocities $v\simeq v_0$ and we thus find $p_{\delta}= (2m_e/3)n_{\delta} v_0^2 = (4/3)n_{\delta} \cE_0$, with $\cE_0=m_ev_0^2/2$.

Our aim is next to derive an expression for the average electron drift in excess of the $E\times B$-drift. We therefore introduce $\udd=\ud-\E\times\B/B^2$. Neglecting inertia, $m\rightarrow0$, we obtain the following momentum equation for electrons within the considered velocity interval:
 \[
0= -e n_{\delta}(\E_{||}+\udd\times\B)-\fd{4\cE_0}{3}\nabla n_{\delta} \quad.
 \]
The current density carried by these electrons is ${\bf J}_{\delta}= -en_{\delta}\udd$ and the momentum balance perpendicular to the magnetic field is then governed by
 \begin{equation}
 \hblabel{eq:momentum}
{\bf J}_{\delta}\times\B=\fd{4\cE_0}{3}\nabla_{\perp} n_{\delta} \quad.
 \end{equation}

An approximate heating rate is readily obtained from the above equation. The average $z$ component of the left hand side may be estimated as
\begin{equation}
\hblabel{eq:JB}
({\bf J}_{\delta}\times\B)_z\simeq \left<J_{\delta y}\right>
\left<B\right>\quad,
\end{equation}
where $\left< ...\right>$ denotes an average over the $z$-direction.
Furthermore, to estimate the right hand side of Eq.~\eq{momentum} we write   $\nabla n_{\delta}\simeq n_{\delta}/D$ such that Eq.~\eq{momentum} approximately yields
\[
\fd{\left< J_{\delta y}\right> }{n_{\delta}}\simeq
\fd{4}{3}\fd{\cE_0}{D \left<B\right>}\quad.
\]
Next, with
$E\scr{rec}\simeq v_A \left<B\right>$ we obtain the following result for the heating rate
 \[
\fd{d\cE}{dt} = \fd{\left< J_{\delta y}\right> E\scr{rec}}{n_{\delta}}\simeq
\fd{4\cE_0 v_A}{3D}\quad.
\]

A comparison with the simulation shows that this expression overpredicts the heating rate by about a factor of two. However, using the numerical profiles to improve the various estimations above (such as  $\nabla n_{\delta}\simeq n_{\delta}/D$ and Eq.~\eq{JB}) we  obtain our final expression for the heating rate consistent with the simulation
  \begin{equation}
  \hblabel{eq:dEdt}
\fd{d\cE}{dt}
\simeq \cE_0 \fd{v_A}{2 D }
\quad.
\end{equation}
Accordingly, electrons trapped in the exhaust double their energy during an Alfv\'enic transit time across the  full width of the exhaust, $2D$.

The energization rate in Eq.~\eq{dEdt} is proportional to the initial energy of the electrons, and is similar to the rate for electrons in  contracting magnetic islands \cite{drake:2006}.
This magnetic island model was introduced to allow for energization in extended reconnection exhausts filled with a bath of magnetic islands. Meanwhile, according to Eq.~\eq{dEdt} electron energization in the reconnection exhaust is effective even without magnetic islands and is not contingent on the development of pressure anisotropy (the latter has been found essential for energization by magnetic island  \cite{drake:2013}). In our analysis, the only requirement for electron energization is that electrons are confined to the reconnection exhaust, which here is aided by the strong $\Phit$ associated with the double layer formation.

We further note that for a magnetized exhaust, our starting point in Eq.~\eq{momentum} can be derived directly by integrating the current carried by the guiding center drifts $\vv_D$ and magnetization current  ${\bf J}_{e\perp}=e\int \vv_D f\,d^3v+\nabla\times{\bf M}$, with the magnetization $e{\bf M}={\bf b}\int \mu f\,d^3v$. Thus, the present analysis is therefore consistent with similar findings obtained using the guiding center model \cite{beresnyak:2014}. In fact, in Appendix A we show  that Eq.~\eq{dEdt} is readily derived using the guiding center model.


\section{Flat-top distributions and superthermal electron energization }

In this Section we explore the heating of electrons in the exhaust. A statistical model is derived which characterizes the energy distribution of the superthermal electrons. While these electrons are mainly energized by perpendicular electric fields, we document how the initial energization and confinement provided by $\Phit$ significantly enhances the effectiveness of the electron energization process.

\subsection{The trapped passing boundaries}

\begin{figure}[h]
\centering
  \includegraphics[width=6.5 cm]{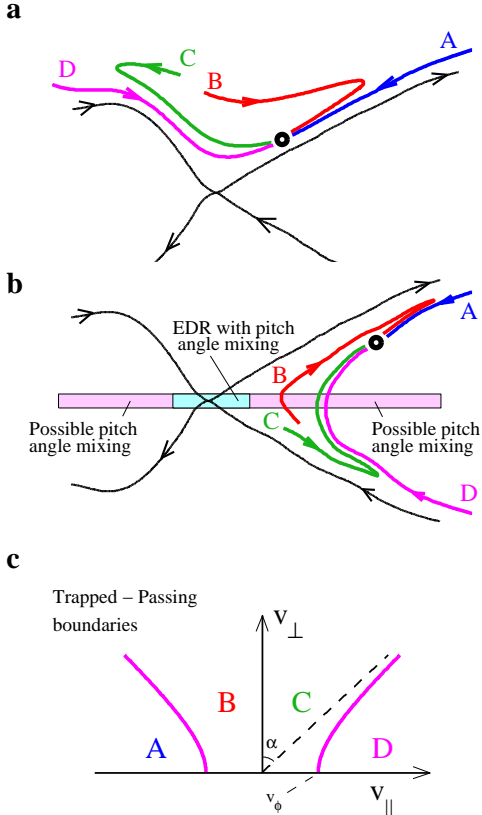}%
  \vspace{-0.5cm}
\caption{a) Classification of electron trajectories reaching a point in the
inflow region. Trajectories $A$ and $D$ are passing while trajectories $B$ and $C$
are trapped. b) Similarly to the inflow region, the trajectories in the exhaust can be classified as passing and trapped. In the exhaust of antiparallel reconnection, pitch angle mixing always occurs at the end of the electron diffusion region. In some cases, pitch angle mixing remains effective far downstream of the reconnection site. c), Regions in the
$(\vz,\vp)$-plane identifying the velocity regions of the four classes of electron trajectories. The
magenta lines represent the trapped/passing boundaries of Eq.~\eq{TPboundary}.
}
\label{fig:orbitdiagB}
\end{figure}

In Figs.~\ref{fig:orbitdiagB}(a,b) we consider different classes of electron trajectories reaching the points highligted.  The trajectories
 passing through the reconnection region along field lines without any reflections are marked by $A$ and $D$, and we denote these as passing. Meanwhile the trajectories marked $B$ and $C$ we denoted as trapped, with electrons bouncing back and forth along field lines, as the field lines convect across  the reconnection region. The four classes of trajectories divide the $(\vz,\vp)$-plane as shown in Fig.~\ref{fig:orbitdiagB}(c).  The trajectories of velocity regions $A$ and $B$  reach the point considered with a negative value of $\vz$, whereas $\vz > 0$ for trajectories of regions $C$ and $D$. The boundary between regions $A$ and $B$ and the boundary between regions $C$ and $D$ we denote the trapped/passing boundaries. In in Fig.~\ref{fig:orbitdiagB}(c) these are shown by the magenta lines.

When the magnetic moment is conserved the perpendicular energy is given by $\cE_{\perp}=\mu B$, implying  that the parallel energy is  $\cE_{\parallel}=\cE-\mu B_{\infty}$. It then follows that the trapped/passing boundaries are  found by solving the equation
\begin{equation}
\hblabel{eq:TPboundary}
\cE_{\parallel\infty}=\cE -e\Phit-\mu B_{\infty}=0\quad,
\end{equation}
which expresses the physical condition that marginally trapped electrons will deplete their parallel energy ($\cE_{\parallel\infty}=0$) as they barely escape along the magnetic field lines away from the reconnection region. We note that for $\vp=0$ the trapped/passing boundaries start at
$|\vz|=v_{\phi}=\sqrt{2e\Phit/m_e}$ and at large $v$ they asymptote to the line in Fig.~\ref{fig:orbitdiagB}(c) characterized by the angle $\alpha$, with  $\cos(\alpha)=B/B_{\infty}$.

\subsection{Size of the loss-cone in velocity space}

For the analysis below it is convenient to introduce $R\scr{loss}$ as a measure of the size of the electron loss-cones, where $4\pi R\scr{loss}$ is the solid angle in velocity space of the loss-cones. Using the expression for the trapped/passing boundary in Eq.~\eq{TPboundary} it is  readily shown that
\begin{equation}
\hblabel{eq:Rloss}
R\scr{loss}=1 - \left[1- \frac{B}{B_{\infty}}\left(1-\frac{e\Phit}{\cE}\right)\right]^{1/2}\quad.
\nonumber
\end{equation}
Introducing the characteristic velocity of the acceleration potential
$v_{\phi}=\sqrt{2e\Phit/m_e}$, an approximate form is obtained for the limit $B\ll B_0$
\begin{equation}
\hblabel{eq:Rloss2}
R\scr{loss}\simeq
\fd{B}{2B_{\infty}}\left(1-\frac{v_{\phi}^2}{v^2}\right)\quad,
\end{equation}
valid for $|\vv|>v_{\phi}$.
For $|\vv|<v_{\phi}$ the confinement is absolute with $R\scr{loss}=0$.

\subsection{Generation of flattop distributions in the pitch angle mixed exhaust}

\begin{figure}[h]
\centering
  \includegraphics[width=6.5 cm]{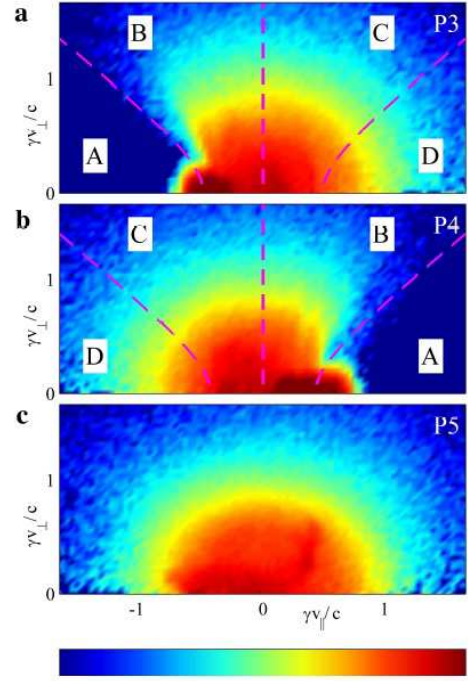}%
\caption{Color contours of the electron distribution function for the points marked P3, P4 and P5 in Fig.~\ref{fig:F1}. For the distributions at P3 and P4 the velocity regions $A$ to $D$ correspond to the distinct trajectories shown in Fig.~\ref{fig:F2}(a) and described in the text. Because of the low values of the magnetic field in the center of the exhaust, pitch angle mixing is effective and the P5 distribution is fully isotropic. For energies $\cE< e\Phit$ the values of $f$ are nearly constant, resembling flattop distribution observed {\sl in situ} in the magnetotail. }
\label{fig:F3}
\end{figure}

The bulk exhaust distributions for the time slice of Fig.~\ref{fig:F1} are displayed in Figs.~\ref{fig:F3}. First, consider the distribution in Fig.~\ref{fig:F3}(a) corresponding to the point marked ``3'' in the panels of Fig.~\ref{fig:F1}. As above, the velocity space (spanned by $\gamma\vz$ and $\gamma\vp$) is divided into regions $A$, $B$, $C$, and $D$ with distinct properties of the associated electron trajectories classified in Fig.~\ref{fig:orbitdiagB}. Again,
the trapped/passing boundaries between regions $A/B$, and regions $C/D$,  are computed using Eq.~\eq{TPboundary}. Trajectories of velocity region $A$ are incoming electrons streaming along field-lines toward the reconnection layer.
The electrons in velocity region $C$ and $D$ are all streaming away from the X-line, but the electrons in region $C$ are trapped  and will  reflect into  region $B$. Meanwhile, the electrons in region $D$ have sufficient parallel energy to escape and will exit the simulation domain at the open boundaries.

The changes in the Fig.~\ref{fig:F3}(a) distribution are dramatic across the $A/B$ boundary calculated using Eq.~\eq{TPboundary}. The region $A$ electrons are well characterized as beams with a parallel energy  $e\Phit$. A small spread is observed about the center of the beams corresponding to the low upstream temperature of these incoming electrons.
The distribution in  Fig.~\ref{fig:F3}(b) is for the point below the midplane marked ``4'' in the panels of Fig.~\ref{fig:F1}. Given the chances in direction of $B_x$, we note how these beams of incoming electrons have opposite signs of $\gamma\vz$ above and below the mid-plane.

The region $A$ beams terminate at the midplane of the simulation domain where, as shown in Fig.~\ref{fig:F1}(d), the magnetic field is weak, $B/B_0<0.1$. This is also evident in the distribution in Fig.~\ref{fig:F3}(c) corresponding to the point marked ``5'' in  Fig.~\ref{fig:F1} on the same field-line as points 3 and 4. The weak magnetic fields causes a complete breakdown of the magnetic moment as an adiabatic invariant and the resulting chaotic electron motion causes the distributions within this layer to become fully isotropic in velocity space.

As the isotropized electrons of velocity region $A$ leave the midplane they populate velocity regions $B$, $C$ and $D$ throughout the exhaust. In principle, these electrons should only add to $f$ at their injection energy $\cE=~e\Phit$. However, distributions in velocity space with negative slope ($df/dv<0$) are unstable and instabilities (including electrons holes) develop in the simulation which are responsible for scattering of the electron energies. These processes and the energization mechanism described by Eq.~\eq{dEdt} drive the distribution function towards  $df/dv\simeq0$, such that for $\cE\leq e\Phit$ flat-top distributions are approached where $f\simeq \mbox{const}$ \citep{nagai:2001,asano:2008,egedal:2012}.

\subsection{Statistical model for superthermal electrons}

\begin{figure}[h]
\centering
  \includegraphics[width=7 cm]{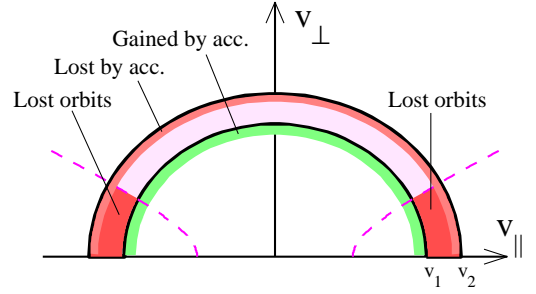}%
\caption{ Regions in velocity space  lost or gained by the velocity shell, $v_1 < |\vv| < v_2$. }
\label{fig:dfdt}
\end{figure}

We now seek to develop a statistical model for the electron energization, which includes the free streaming losses for electrons at energies above $e\Phit$. Given the pitch angle mixing in the exhaust, the phase space density is assumed to be fully isotropic $f(\vv)=f(v)$. To obtain an evolution equation for $f(v)$ we consider the phase-space volume bounded by $v_1$ and $v_2$ in Fig.~\ref{fig:dfdt}. This spherical shell in velocity space has a width $dv=v_2-v_1$ and a volume $4\pi v_1^2dv$.  The number of electrons in this volume changes because of three effects. 1)
Electrons within the losscone of size  $4\pi v_1^2dvR\scr{loss}$ are all lost within half a bounce time $\tau_b/2$.
2)
Electrons with velocities just below $v_2$ will be  lost by acceleration across the $v=v_2$ boundary. 3) Similarly,  electrons with velocities just below $v_1$ will be accelerated into the volume. Considering a time interval $dt$, particle conservations may then be expressed as
\begin{eqnarray}
\hblabel{eq:mass1}
4\pi v^2\hspace{-0.2cm} &dv&\hspace{-0.2cm} \left. \dd{f}{t} dt \right|_{v=v_1}=
 -  \left. 4\pi v^2dv\,f\,R\scr{loss}
\fd{dt}{\tau_b/2}\right|_{v=v_1}   \nonumber \\[1ex]
&-&4\pi \left. v^2 \dd{v}{t} dt f \right|_{v=v_2}
+4\pi \left. v^2 \dd{v}{t} dt f \right|_{v=v_1}
\,\,. \nonumber
\end{eqnarray}
It is clear that the last two terms in the equations above can be rewritten as $dv$ times a differential. Dividing through by $4\pi\, dv \,dt$ we then obtain.
\begin{equation}
\hblabel{eq:mass2}
v^2  \dd{f}{t}= - \fd{2v^2}{\tau_b}f\, R\scr{loss}
-\dd{ }{v}\left( v^2\dd{v}{t} f\right)
\quad.
\end{equation}

To proceed requires estimations of $\tau_b$ and $\ddt{v}{t}$ as functions of $v$.
From the trajectory on the right hand side of Fig.~\ref{fig:pltOrbits}(c) we conclude that the characteristic orbit length is about $2D$, where (as above) $D$ is the half width of the reconnection exhaust. Then, the orbit bounce time is approximately $\tau_b\simeq 2D/v$.
Furthermore, by using $d\cE/dt=mv\,dv/dt$
it follows from Eq.~\eq{dEdt} that
$\ddt{v}{t}\simeq v \,v_A /(4D)$. Inserting this (and $\tau_b=2D/v$) into Eq.~\eq{mass2} we obtain
\begin{equation}
\hblabel{eq:mass3}
4l\,v^2  \dd{f}{t} +v_A \dd{ }{v}\left( v^3 f\right)= - 4 v^3f\, R\scr{loss}
\quad.
\end{equation}
With the expression in Eq.~\eq{Rloss2} for $R\scr{loss}(v)$ a steady state solution ($\ddt{f}{t}=0$) is then
\begin{equation}
\hblabel{eq:fexp}
f\scr{model}\propto\fd{1}{v^3}\exp\left(-\fd{2B}{B_0}\fd{(v-v_{\phi})^2}{v\,v_A}\right)\quad,
\end{equation}
applicable for  energies above $e\Phit$ (or $v>v_{\phi}$).

\subsection{Comparison of model for superthermal electrons to kinetic simulation data}

\begin{figure}[h]
\centering
  \includegraphics[width=8 cm]{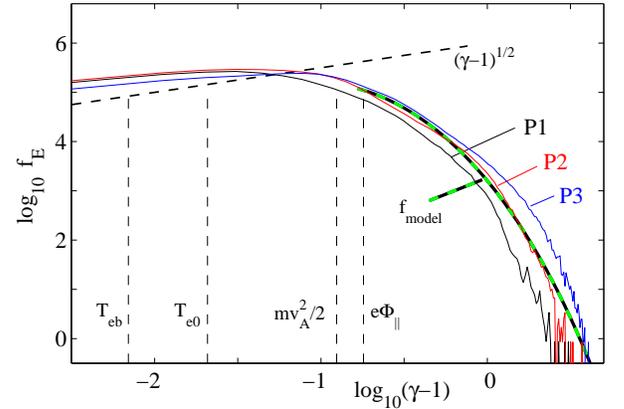}%
\caption{ Electron energy distribution $f_{E}$ for the points marked P1 -- P3 in Fig.~\ref{fig:F1}. For $\cE<e\Phit$ the flattop part of the distribution is observed with $f_{E}\propto (\gamma-1)^{1/2}$. The dashed green/black line is $f\scr{model}$ in Eq.~\eq{fexp} applicable for  $\cE> e\Phit$, evaluated with $v_A/c=0.08$, $e\Phit/mc^2=0.22$ and $B/B_{\infty}=0.3$. }
\label{fig:F4}
\end{figure}

To compare Eq.~\eq{fexp}  to the kinetic simulation we use the energy distribution $f_{E}$, here defined as a function of the relativistic kinetic energy $\cE=m_0c^2(\gamma-1)$. Using $n_e=\int f_E d(\gamma-1)= \int f d^3v$, at non-relativistic energies we have $f_E\propto (\gamma-1)^{1/2}\, f$. Fig.~\ref{fig:F4} shows $f_E$ for the points marked 1 -- 3 in Fig.~\ref{fig:F1}, with the most energetic distribution corresponding to point 3, the furthest point away from the x-line. For $m_0c^2(\gamma-1)< e\Phit$ we approximately have $f_E\propto(\gamma-1)^{1/2}$ corresponding to $f\simeq \mbox{const}$ for the near flat-top part of the distributions. For energies above $e\Phit$ the form in Eq.~\eq{fexp} represents a good  approximation to the simulation data.

The distributions in Fig.~\ref{fig:F4} demonstrate that significant electron energization is possible in an open exhaust of a single X-line  reconnection configuration.
We find that not only does $\Phit$ provide the electrons with an initial energy boost as they enter the reconnection region, the structure of $E_{\parallel}$ also reduces the free-streaming along field lines in the reconnection exhaust. This permits the  accumulation of energetic electrons, heated mainly by perpendicular electric fields ($E_{\perp}$) during their repeated bounce-motion across the exhaust. Energization thereby becomes effective throughout the reconnection exhaust at much larger scales than the kinetic length scales of the electrons and ions.

\section{Comparisons of numerical results with  spacecraft observations}

\subsection{Electron bulk energization  during magnetotail reconnection}

Several decades of {\sl in situ} spacecraft observations show
that strong kinetic effects are present during magnetic reconnection in the Earth's magnetosphere \cite{lin:1995,nagai:2001,oieroset:2002}.
Particularly relevant to the role of $\Phit$  are the electron distributions documented by Nagai {\sl et al.}, which, similarly to Fig.~\ref{fig:F3}(a,b), reveal the presence of cold beams directed towards the X-line while energized electrons move away from the reconnection region (see Fig.~4 in Ref.~\cite{nagai:2001}).
The importance of these electron beams is also emphasized by measurements
taken by the Geotail and the Cluster spacecraft
\cite{nagai:2001,asano:2008,chen:2008jgr}. Within
the reconnection outflow the electrons often have a characteristic
isotropic flat-top distribution, where the phase space density,
$f$, of the electrons is nearly constant from thermal energies
(tens to hundreds of eV) up to several keV. As detailed by Egedal {\sl et al.}~\cite{egedal:2012}, the distributions of Fig.~\ref{fig:F3} resemble closely distributions observed during reconnection events in the magnetotial. Furthermore, the extensive study of flat-top distributions by Asano {\sl et al.}~provides the spatial structure of the magnetotail exhausts where these flat-top distributions are observed (see Fig. 22 in Ref.~\cite{asano:2008}), which resemble the structure  of $\Phit$ in Fig.~\ref{fig:F1}).

The kinetic simulation results, the scaling law for $\Phit$ in Eq.~\eq{ephi}, and threshold  in Eq.~\eq{adia3} for non-adiabatic electron behaviour all suggest that $\beta_{e\infty}$ is an important parameter for the dynamics of the electrons during reconnection. To explore the validity of these results to reconnection in the magnetotail we study the inflow values of $\beta_{e}$ as well as $\Phit$ inferred from a number of reconnection events encountered by the Cluster mission.

The four Cluster satellites are moving together in coordinated polar orbits around the Earth, their internal separation changing over the years of operation (year 2000 to present day). The data analyzed here is obtained through the Cluster Active Archive. The magnetic field data is  provided by the Flux-Gate Magnetometer (FGM) experiment \cite{balogh:2001} and the ion plasma data by the Cluster Ion Spectrometry (CIS) experiment \citep{reme:2001}. The electron data are from the Plasma Electron and Current Experiment (PEACE) \cite{johnstone:1997}. In the analysis, only the PEACE data points with values above the background electron flux count that were also flagged with quality number 3 or 4 (good for publication) were used. Both the HEEA (high energy sensor) and the LEEA (low energy sensor) data were included. To avoid photo electron measurements we consider only electrons at energy levels higher than 70 eV.

The data selected for our analysis were recorded during encounters with 21 separate magnetic reconnection events in the Earth's magnetotail at about 18 Re. The reconnection events are listed in Ref.~\cite{borg:2012a} along with the signatures of magnetic reconnection, as observed by spacecraft flying through the reconnection region on a path parallel to the magnetotail neutral sheet. During all of the encounters, some or all of the spacecraft observed both reconnection outflow regions. The inflow regions were observed during some of the encounters. The signatures of both inflow and outflow regions are described in  Ref.~\cite{borg:2012a}. It must be noted that while the outflow region is easily identified in the data, the inflow region is less distinct and this characteristic may introduce errors.

\begin{figure}[h]
\centering
  \includegraphics[width=9 cm]{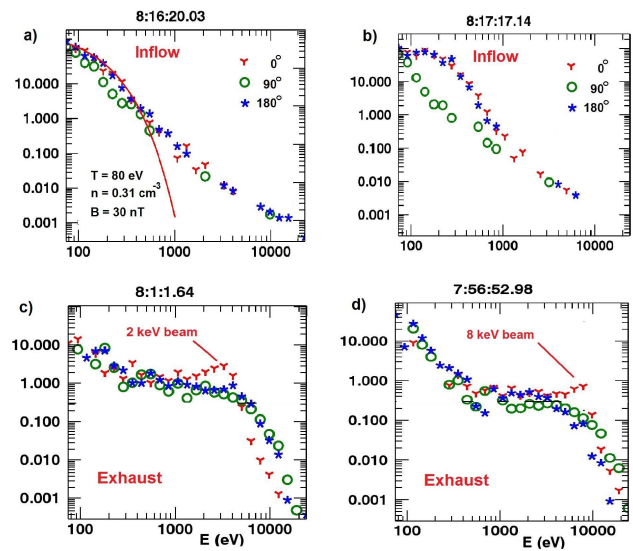}%
\caption{Characteristic electron distribution functions recorded by the Cluster Mission during a reconnection event on August 21, 2002. }
\label{fig:Cluster}
\end{figure}

 To determine the upstream electron characteristics for  the observed magnetic reconnection encounters
we considered data recoded by spacecraft during excursions from the outflow regions into areas of lower plasma density and stronger $B_x$ magnetic field.  Still, these excursions did not bring the spacecraft into the magnetotail lobe. As an example of the data applied in this analysis, Fig.~\ref{fig:Cluster} shows electron phase-space distributions recorded by $Cluster$ 4 during a reconnection event encountered on August 21, 2002. The distribution in Fig.~\ref{fig:Cluster}(a) is representative of the upstream reconnection region, and it's thermal component is well described by a Maxwellian fit yielding inflow parameters of $T_e\simeq80$ eV and  $n_e\simeq0.31$ cm$^{-3}$. At the magnetic field of $B\simeq30$ nT, the normalized electron pressure for the inflow can then be estimated as $\beta_{e\infty}\simeq 0.011$.

 The distribution in Fig.~\ref{fig:Cluster}(b) is also for a location in the inflow region where the observed electron anisotropy develops in agreement with the kinetic model in Refs.~\cite{egedal:2008jgr,egedal:2013}. Meanwhile, the distributions in Figs.~\ref{fig:Cluster}(c,d) are observed in the exhaust and resemble the numerical distributions of Figs.~\ref{fig:F3}(a,b). Given this resemblance, a local value of $e\Phit$ can be identified as the energy of the incoming electron beams and/or the energy at which the flattop part of the distributions terminate. By considering all the measured distributions available, the maximal value of $\Phit$ for this event was estimated to be $\Phit\simeq$ 9 keV, corresponding to a normalized potential $e\Phit/T_e\simeq 110$ similar to the numerical values obtained above.

To obtain the lobe plasma beta values we applied the electron and magnetic field data for the observations of the magnetotail lobe closest in time to the reconnection outflow observations. In table \ref{betas} we provide the values of $\Phit$, $\beta\scr{lobe}$ as well as the inflow values of $n_e$, $T_e$, $B$, and $\beta\scr{inflow}(=\beta_{\infty})$ estimated for the reconnection events. Out of the 21 events analyzed, we obtained inflow and lobe plasma beta values for 18 of them  using data from the FGM and the PEACE instruments. Compared to the inflow values $\beta\scr{inflow}$ characterizing the plasma feeding the reconnection region, the lobe values $\beta\scr{lobe}$ were typically lower
by an order of magnitude.

\begin{figure}[h]
\centering
  \includegraphics[width=7.4 cm]{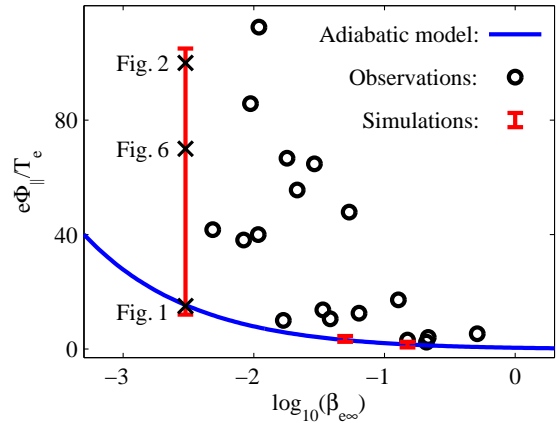}%
\caption{Values of $e\Phit/T_e$ as a function of $\beta_{e\infty}$. The black circles are calculated based on the values in Table~\ref{betas} inferred from Cluster spacecraft data. The red symbols represent values of kinetic simulations. The values of $e\Phit/T_e$  for $\beta_{e\infty}=0.05$ and 0.15 are from simulations presented in Ref.~\cite{le:2010grl}. The blue line is the adiabatic model in Eq.~\eq{ephi}. This model  does not include the effects on $\Phit$ caused by electron holes and double layers, and thus represents a lower theoretical bound for $e\Phit/T_e$. }
\label{fig:data}
\end{figure}

The results from the described analysis of the Cluster data are summarized in Fig.~\ref{fig:data}, which provides the values of $e\Phit/T_e$ as a function of $\beta_{e\infty}$. For comparison, the red symbols represent the range of $e\Phit/T_e$ recorded  in the numerical simulations (with values from particular figures above indicated). In addition, the blue line is $e\Phit/T_e$ of Eq.~\eq{ephi}, which yields  a lower bound for the observations and simulations.

The range of $e\Phit/T_e$ from the simulation presented here (with $\beta_{e\infty}=0.003$) is in good agreement with the spacecraft observations. However, the spacecraft data suggest that the transition to the regime influenced by electron holes and double layers (large values of $e\Phit/T_e$) occurs at  $\beta_{e\infty}\simeq 0.1$. In contrast, all simulations we have studied with $\beta_{e\infty}\simeq 0.1$ are well described by Eq.~\eq{ephi} (the blue line) \cite{le:2010grl}. Nevertheless, the threshold for the transition to non-adiabatic inflow electron, $\beta_{e\infty}<0.02$ in Eq.~\eq{adia3}, is in reasonable agreement with the spacecraft observations.

\begin{table}
\caption{Plasma beta and guide field values for magnetic reconnection inflow regions and the lobe.}
\centering
\begin{tabular}{l c l r c l l l}
\hline
 Date & $\Phit$/kV & $\beta\scr{lobe}$ & $T_e$/eV & \qquad& $n_e/$cm$^{-3}$ & $B/$nT &$\beta\scr{inflow}$ \\
\hline
2001 08 22 & 8& 0.003 & 210 & &0.07 & 26 & 0.008 \\
2001 09 10 & 0.8 & 0.003 & 340 & & 0.36 & 15 & 0.21  \\
2001 09 12 & 5 & 0.001 & 90 & &0.25 & 20 & 0.026 \\
2001 10 01 & 10& 0.008 & 150 & & 0.28 & 30 & 0.018 \\
2001 10 08 & 2 & 0.004 & 500& & 0.45 & 20 & 0.22  \\
2001 10 11 & 2 & 0.03  & 650& & 0.14 & 15 & 0.15 \\
2002 08 21 & 9 & 0.003 & 80 & & 0.32 & 30 & 0.011 \\
2002 08 28 & 2.1 & 0.0003 &200 & &0.2 &  20 & 0.038\\
2002 09 13 & 3 & 0.001 & 220 & &0.23 & 24 & 0.034 \\
2002 09 18 & 11 & 0.0009 & 230 & & 0.24 & 20 & 0.054  \\
2002 10 02 & 11 & 0.002 & 170 & & 0.064 & 12 & 0.030 \\
2003 08 17 & 5  & 0.0003 & 120 & & 0.17 & 40 & 0.0048  \\
2003 08 24 & 6  & 0.0006 & 350& & 0.38 &  20  & 0.13 \\
2003 09 19 & 8  & 0.002 & 1500& & 0.2 &  15 & 0.51  \\
2003 10 04 & 6  & 0.003 & 70 & & 0.14  & 20 & 0.0094 \\
2003 10 09 & 5  & 0.002  & 400& & 0.12 & 17 & 0.064 \\
2004 09 14 & 6  & 0.002 & 150& & 0.075 & 20 & 0.011\\
2005 09 26 & 1  & 0.001 &100 & & 0.70 &  40 & 0.017  \\
\hline
\end{tabular}
\label{betas}
\end{table}

\subsection{Superthermal electrons in the magnetotail}

\begin{figure}[h]
\centering
  \includegraphics[width=7.4 cm]{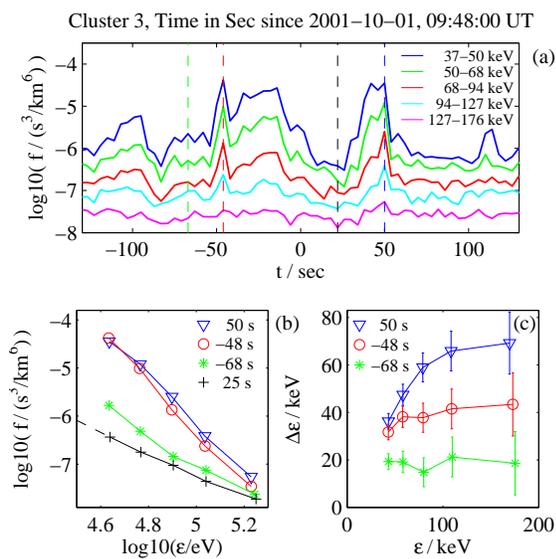}%
\caption{Electron phase-space densities from the RAPID
measurement on Cluster 3. (b) Electron distribution observed
at separate time points. (c) Spectra of $\Delta\cE$ for three selected times. The observed spectra of $\Delta\cE$, independent of $\cE$ for large $\cE$, are consistent with Eq.~\eq{dEdt} when taking into account the finite extent of the systems in the magnetotail $y$-direction.}
\label{fig:super}
\end{figure}

In a previous study of magnetotail observations we found that superthermal electrons often acquire a constant energy-gain, $\Delta\cE$, independent of their initial energy \cite{egedal:2010grl}. As an example, in Fig.~\ref{fig:super} we provide measurements of superthermal electrons recorded by the Cluster Mission during the much studied October 1, 2001 reconnection event \cite{wygant:2005}. Time series measurements by the RAPID instrument are given in a). For the selected time points for which the distributions are shown  in b), we apply a Liouville mapping technique to obtain the energy gains  shown in c). Again, more details of this analysis are giving in  Ref.~\cite{egedal:2010grl}.

For all the events analyzed it is found that, for the most energitic electrons, $\Delta\cE$ is constant, independent of the initial energy. In Ref.~\cite{egedal:2010grl} we then concluded that the energization of the superthermal electrons is set by $e\Phit$, but our understanding of this has changed and is now very different. From the above analysis it is clear that it is the bulk-energization (including flat-top distributions) that is controlled by $e\Phit$, and the observed superthermal electrons energization is  better described by Eq.~\eq{dEdt}, allowing for energization to energies  much larger than $e\Phit$.
However, because of the initial hard power-law spectrum $f(v)\propto v^{-3}$ of incoming lobe electrons and because $\Delta\cE$ in Eq.~\eq{dEdt} is proportional to the initial energy, the present 2D model predicts superthermal electrons in excess of the observations. This discrepancy can be explained by the finite $y$-extent of the magnetotail. In a realistic application, the effectiveness of the present model needs to be limited, as the maximum energization level is set by the potential drop along the extent of the reconnection x-line. The constant levels of $\Delta\cE$ deduced in the analysis of Ref.~\cite{egedal:2010grl} are then consistent with our superthermal heating mechanism in combination with a finite dusk-dawn (or for observations: dusk - spacecraft) potential drop during magnetotial reconnection.

\subsection{Solar flares}

Compelling observational evidence exists for confinement of energetic electrons by parallel electric fields during solar flare events. This evidence is summarized in papers by Li.~{\sl et al.}, \cite{li:2012,li:2013,li:2014} who (as mentioned above) explored the role of double layers for reducing the free streaming losses of reconnection energized electrons along the field lines. The evidence includes the formation of energetic electron populations detected in the vicinity of the expected loop-top reconnection sites. The lifetime of these energetic populations are inferred to be much longer than the thermal electron transit time \cite{krucker:2010}, suggesting that parallel electric fields must be important to reduce free streaming losses along magnetic field lines.

\begin{figure}[h]
\centering
  \includegraphics[width=9 cm]{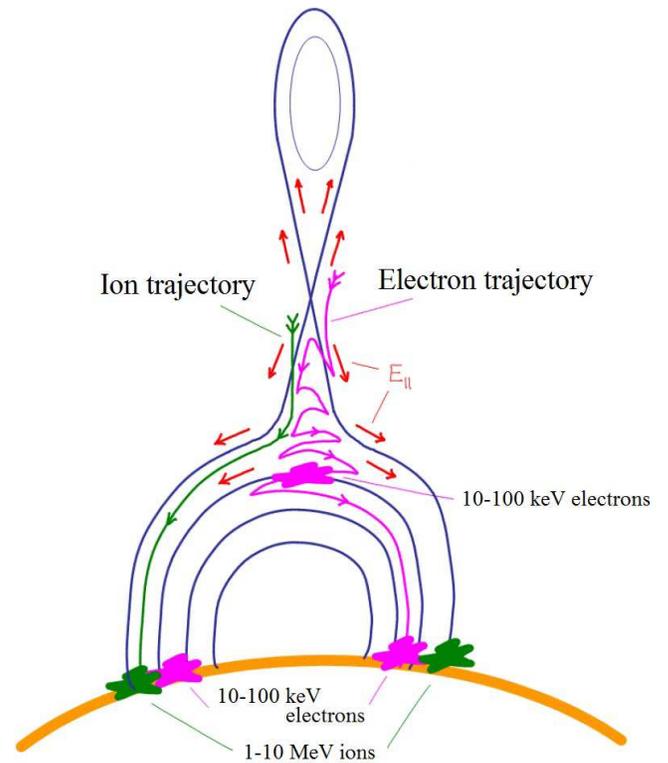}%
\caption{Schematic model of a solar flare }
\label{fig:flare}
\end{figure}

Based on the Masuda flare model \cite{masuda:1994}, Fig.~\ref{fig:flare} provides a schematic illustration of the electric and magnetic geometry suggested by the observations and required for confinement of electrons by parallel electric fields. Again, the analysis by Li.~{\sl et al.} and also earlier authors (see references in Ref.~\cite{li:2012}) have already suggested that electric fields in double layers may help confine the energized electrons. In addition, the present analysis suggests that the parallel electric fields are not only important for confining the electrons but are an integral component to the overall energization processes. For example, the observed and remarkably efficient energization of the bulk electrons from about 1keV up to 10-100keV \cite{krucker:2010} is consistent with the development of an acceleration potential with a magnitude of $e\Phit/T_e=100$ seen in the magnetotail and in our simulation. In addition, the most energetic electrons reaching the MeV energy range can also be accounted for by Eq.~\eq{fexp}. In fact, for energies just above $e\Phit$, Eq.~\eq{fexp} yields a hard powerlow spectrum $f(v)\propto v^{-3}$ for about an order of magnitude in energy, sufficient to explain the most energetic electrons in the solar flare observations.
We also note that the $\beta_{e\infty}$-threshold ($\beta_{e\infty}<0.02$)   for the non-adiabatic electron dynamics to occur is easily satisfied in solar flare events \cite{fletcher:2011}.

In our model, the electrons are only energized by $\Phit$ as they enter the reconnection exhaust. Meanwhile, ions will be accelerated away from the reconnection region by these parallel electric fields. The path of the ions will not coincide with instantaneous magnetic field lines, and a fraction of the ions are likely to endure acceleration by $E_{\parallel}$ to much larger values than $e\Phit$ during their transit  through the reconnection exhaust. This may account for the energization of ions in the range of 10 MeV recorded by the RHESSI spacecraft \cite{lin:2003}. Of course, it must be noted that our numerical simulation covers a domain size of about 300$d_i$, whereas the size of a solar flare is about $10^7d_i$. Furthermore, solar flares are believed to include a guide magnetic field of order unity, whereas the present simulation is for anti-parallel reconnection. Nevertheless, we expect that the described general processes leading to non-adiabatic parallel electron dynamics and strong electron energization will also be  applicable to reconnection in very large systems.

\section{Conclusions}
The heating by ${\bf E}\cdot \vv_{D}$ has been found generic to plasma flows driven by magnetic tension \cite{beresnyak:2014} and is common to a range of models for electron energization. An example is Drake's Fermi acceleration model  considering a bath of reconnecting magnetic islands \cite{drake:2005}. Likewise the term ${\bf E}\cdot \vv_{D}$ is also  responsible for driving powerlaw distributions in highly relativistic pair plasmas \cite{guo:2014}. A third example is the energization documented while propagating test-particles  through the fields of Ideal-MHD simulations \cite{birn:2013}, where confinement may occur when electrons bounce between dipolarization fronts and the stronger magnetic fields close to Earth \cite{runov:2011,birn:2013}.
However, Ideal-MHD assumes $E_{\parallel}=0$ and the described effects of localized electron trapping and the initial energization by $E_{\parallel}$ are therefore omitted.

The key new feature presented here is the importance of the magnetic field aligned electric fields  $E_{\parallel}$ for initial energization of the electrons. Because the subsequent energization by $E_{\perp}$ is propotional to $\cE$, the initial energy boost that electrons acquire from $e\Phit$
largely determines the overall efficiency of the energization process.
The large amplitude structure of $\Phit$ also limits the
free streaming particle losses along magnetic field lines, which is essential for confining the electrons and shaping  the spectra of their energy distribution. The heating is effective in the full exhaust of a single X-line configuration, and does not require pressure anisotropy to develop.

The large values of $E_{\parallel}$ are observed within density cavities and develop in association with the  strong double layers. A threshold, $\beta_{e\infty} < 0.02$ is derived for the strong double layer formation, suggesting that these are likely to be present during solar flare events. Indeed, the level of electron energization observed in the simulation is consistent with that observed during solar flares. Furthermore, we show the details of the electron distribution functions described in our analysis are in agreement with {\sl in situ} spacecraft measurements obtained over the last decade  during reconnection events in the Earth's magnetotail.

\section*{Acknowledgments}
The work at UW-Madison
was funded in part by NASA grant NNX14AC68G. The numerical simulation work  was supported by the NASA Heliophysics Theory Program at LANL. Initial simulations were carried out using LANL institutional computing resources and the Pleiades computer at NASA, while the final simulation was
carried out on Kraken with an allocation of advanced computing resources provided by the National Science Foundation at the National Institute for Computational Sciences (http://www.nics.tennessee.edu/).

\section*{Appendix A: Use of the guiding center model for estimating the electron energization rate}
\label{guide}

In this appendix we first discuss why the guiding center model can provide an accurate prediction of the electron heating rate for the pitch angle mixed exhaust. We then estimate the heating rate based on this model and recover result given above in Eq.~\eq{dEdt}. We thereby validate the above estimate and highlight the similarity (and differences) between the results in the present paper compared to those of earlier works using the guiding center model \cite{dahlin:2014}.

From Fig.~\ref{fig:F2}(b) above it is clear that the main heating source of the electrons is the perpendicular electric fields, and thus, the local heating results from the term $en_e{\bf V}_{e\perp}\cdot\E$, where ${\bf V}_{e\perp}$ is the electron fluid velocity also appearing in the electron momentum equation:
\begin{equation}
\hblabel{eq:Emomentum}
\E+ {\bf V}_{e}\times\B=-\nabla\cdot\bar{\bar{{\bf P}}}_e/(en_e)
\quad.
\end{equation}
 Due to the low values of $B$ in the exhaust, the magnetic moments of the electrons are not conserved. The particle motion therefore becomes chaotic and all pitch angle information is lost as electrons pass through the midplane. We are therefore only interested in the average electron behavior as a function of energy. It is clear that the guiding center  approximation with the  drifts  $\vv_D=\vv_k+\vv_{\nabla B}$ does not formally apply (where $\vv_k=m_e\vz^2\B\times\kappa/(eB^2)$ with $\kappa= {\bf b}\cdot\nabla{\bf b}$ is the curvature drift and  $\vv_{\nabla B}=m_ev_{\perp}^2\B\times\nabla B/(2eB^2)$ is the gradient-B drift). Nevertheless, by direct evaluation of the expression $en_e
{\bf V}_{e\perp}=e\int \vv_D f\,d^3v+\nabla\times{\bf M}$, with the magnetization $e{\bf M}={\bf b}\int \mu f\,d^3v$,  it is well known that  ${\bf V}_{e\perp}$ in Eq.~\eq{Emomentum} is recovered for the relevant limit where $\bar{\bar{{\bf P}}}_e$ has no off diagonal stress \cite{hazeltine:1992}.
Therefore, although the guiding center approximation does not account for the chaotic motion of the individual electrons, it does accurately predict  the total fluid drift. Furthermore, the predicted contribution to ${\bf V}_{e\perp}$ from particles within a given energy interval is also accurate. The guiding center approximation thus allows  us to estimate the {\sl average} electron drift (and associated energization) as a function of  energy.

To derive an expression for $d\cE/dt$ using the guiding center model,  we note that previous studies in similar geometries have found that curvature drifts dominate the particle motion in the direction of the reconnection electric field \cite{dahlin:2014}. Therefore, we here  estimate the energy changes over a single bounce orbit caused by the curvature drift, $\Delta\cE=e\int_0^{\tau_b}\vv_k\cdot\E\scr{rec}dt$. Using $|\kappa|=1/R_c$ where $R_c$ is the radius of curvature for the magnetic field line and $\tau_b\simeq\pi R_c/\vz$ we obtain $\Delta\cE\simeq E\scr{rec}\pi m_e|\vz|/B\simeq E\scr{rec}mv/B$, as the  average energization per bounce.  Using $E\scr{rec}\simeq v_AB_0/10$ we obtain
\begin{equation}
\fd{d\cE}{dt}=\fd{\Delta\cE}{\tau_b}\simeq\fd{v_AB_0}{10} \fd{mv}{B {\tau_b}}\simeq  \fd{v_A}{l\scr{orb}} mv^2 \quad,
\nonumber
\end{equation}
where we have used $B\simeq B_0/10$.
Thus, we find that
\begin{equation}
\hblabel{eq:dEdt2}
\fd{d\cE}{dt}\simeq \fd{2v_A}{l\scr{orb}} \cE\quad.
\end{equation}
This order of magnitude estimate of the heating rate is consistent with the result in Sec.~\ref{heating}. In fact, taking the orbit length as four times the half exhaust width, $l\scr{orb}=4D$, we recover the expression in Eq.~\eq{dEdt}.

\bibliographystyle{ieeetr}


\begin{thebibliography}{10}

\bibitem{krucker:2010}
S.~{Krucker}, H.~S. {Hudson}, L.~{Glesener}, S.~M. {White}, S.~{Masuda}, J.-P.
  {Wuelser}, and R.~P. {Lin}, ``{Measurements of the Coronal Acceleration
  Region of a Solar Flare},'' {\em \apj}, vol.~714, pp.~1108--1119, May 2010.

\bibitem{oieroset:2001}
M.~{\O}ieroset, T.~Phan, M.~Fujimoto, R.~P. Lin, and R.~P. Lepping, ``In situ
  detection of collisionless reconnection in the earth's magnetotail,'' {\em
  Nature}, vol.~412, pp.~414--417, JUL 26 2001.

\bibitem{drake:2006}
J.~F. Drake, M.~Swisdak, H.~Che, and M.~A. Shay, ``Electron acceleration from
  contracting magnetic islands during reconnection,'' {\em Nature}, vol.~443,
  pp.~553--556, OCT 5 2006.

\bibitem{oka:2010}
M.~Oka, T.~D. Phan, S.~Krucker, M.~Fujimoto, and I.~Shinohara, ``{ELECTRON
  ACCELERATION BY MULTI-ISLAND COALESCENCE},'' {\em {ASTROPHYSICAL JOURNAL}},
  vol.~{714}, pp.~{915--926}, {MAY 1} {2010}.

\bibitem{hoshino:2012}
M.~Hoshino, ``{Stochastic Particle Acceleration in Multiple Magnetic Islands
  during Reconnection},'' {\em Phys. Rev. Lett.}, vol.~{108}, {MAR 28} {2012}.

\bibitem{drake:2013}
J.~F. Drake, M.~Swisdak, and R.~Fermo, ``{THE POWER-LAW SPECTRA OF ENERGETIC
  PARTICLES DURING MULTI-ISLAND MAGNETIC RECONNECTION},'' {\em {ASTROPHYSICAL
  JOURNAL LETTERS}}, vol.~{763}, {JAN 20} {2013}.

\bibitem{chen:2008}
L.~J. Chen, A.~Bhattacharjee, P.~A. Puhl-Quinn, H.~Yang, N.~Bessho, S.~Imada,
  S.~Muehlbachler, P.~W. Daly, B.~Lefebvre, Y.~Khotyaintsev, A.~Vaivads,
  A.~Fazakerley, and E.~Georgescu, ``Observation of energetic electrons within
  magnetic islands,'' {\em Nature Physics}, vol.~4, pp.~19--23, JAN 2008.

\bibitem{oieroset:2012}
M.~Oieroset, T.~D. Phan, J.~P. Eastwood, M.~Fujimoto, W.~Daughton, M.~A. Shay,
  V.~Angelopoulos, F.~S. Mozer, J.~P. McFadden, D.~E. Larson, and K.~H.
  Glassmeier, ``{Direct Evidence for a Three-Dimensional Magnetic Flux Rope
  Flanked by Two Active Magnetic Reconnection X Lines at Earth's
  Magnetopause},'' {\em Phys. Rev. Lett.}, vol.~{107}, {OCT 13} {2011}.

\bibitem{block:1978}
L.~BLOCK, ``Double-layer review,'' {\em ASTROPHYSICS AND SPACE SCIENCE},
  vol.~{55}, no.~{1}, pp.~{59--83}, {1978}.

\bibitem{singh:1987}
N.~SINGH, H.~THIEMANN, and R.~SCHUNK, ``Electric-fields and double-layers in
  plasmas,'' {\em LASER AND PARTICLE BEAMS}, vol.~{5}, pp.~{233--255}, {MAY}
  {1987}.

\bibitem{raadu:1988}
M.~RAADU and J.~RASMUSSEN, ``Dynamical aspects of electrostatic
  double-layers,'' {\em ASTROPHYSICS AND SPACE SCIENCE}, vol.~{144},
  pp.~{43--71}, {MAY} {1988}.

\bibitem{chen:2008jgr}
L.~J. Chen, N.~Bessho, B.~Lefebvre, H.~Vaith, A.~Fazakerley, A.~Bhattacharjee,
  P.~A. Puhl-Quinn, A.~Runov, Y.~Khotyaintsev, A.~Vaivads, E.~Georgescu, and
  R.~Torbert, ``{Evidence of an extended electron current sheet and its
  neighboring magnetic island during magnetotail reconnection},'' {\em J.
  Geophys. Res.}, vol.~{113}, {DEC 19} {2008}.

\bibitem{egedal:2008jgr}
J.~Egedal, W.~Fox, N.~Katz, M.~Porkolab, M.~{\O}ieroset, R.~P. Lin,
  W.~Daughton, and D.~J. F., ``Evidence and theory for trapped electrons in
  guide field magnetotail reconnection,'' {\em J. Geophys. Res.}, vol.~113,
  p.~A12207, MAR 25 2008.

\bibitem{egedal:2012}
J.~Egedal, W.~Daughton, and A.~Le, ``Large-scale electron acceleration by
  parallel electric fields during magnetic reconnection,'' {\em Nature
  Physics}, vol.~8, pp.~321--324, APR 2012.

\bibitem{fletcher:2011}
L.~Fletcher, B.~R. Dennis, H.~S. Hudson, S.~Krucker, K.~Phillips, A.~Veronig,
  M.~Battaglia, L.~Bone, A.~Caspi, Q.~Chen, P.~Gallagher, P.~T. Grigis, H.~Ji,
  W.~Liu, R.~O. Milligan, and M.~Temmer, ``An observational overview of solar
  flares,'' {\em Space Science Reviews}, vol.~159, pp.~19--106, SEP 2011.

\bibitem{li:2012}
T.~C. Li, J.~F. Drake, and M.~Swisdak, ``Supression of energetic electron
  transport in flares by double layers,'' {\em Astrophys. J.}, vol.~757, SEP 20
  2012.

\bibitem{le:2009}
A.~Le, J.~Egedal, W.~Daughton, W.~Fox, and N.~Katz, ``{Equations of State for
  Collisionless Guide-Field Reconnection},'' {\em Phys. Rev. Lett.},
  vol.~{102}, {FEB 27} {2009}.

\bibitem{egedal:2013}
J.~Egedal, A.~Le, and W.~Daughton, ``A review of pressure anisotropy caused by
  electron trapping in collisionless plasma, and its implications for magnetic
  reconnection,'' {\em Phys. Plasmas}, vol.~20, JUNE 2013.

\bibitem{bowers:2009}
K.~Bowers, B.~Albright, L.~Yin, W.~Daughton, V.~Roytershteyn, B.~Bergen, and
  T.~Kwan, ``{Advances in petascale kinetic plasma simulation with VPIC and
  Roadrunner},'' {\em {Journal of Physics: Conference Series}}, vol.~{180},
  p.~{012055 (10 pp.)}, {2009} {2009}.

\bibitem{egedal:2009pop}
J.~Egedal, W.~Daughton, J.~F. Drake, N.~Katz, and A.~Le, ``{Formation of a
  localized acceleration potential during magnetic reconnection with a guide
  field},'' {\em Phys. Plasmas}, vol.~{16}, {MAY} {2009}.

\bibitem{chew:1956}
G.~F. Chew, M.~L. Goldberger, and F.~E. Low, ``The boltzmann equation and the
  one-fluid hydromagnetic equations in the absence of particle collisions,''
  {\em Proc.~Royal Soc.~A}, vol.~112, p.~236, 1956.

\bibitem{le:2010grl}
A.~Le, J.~Egedal, W.~Daughton, J.~F. Drake, W.~Fox, and N.~Katz, ``{Magnitude
  of the Hall fields during magnetic reconnection},'' {\em Geophy. Res. Lett.},
  vol.~{37}, {FEB 11} {2010}.

\bibitem{ng:2011}
J.~Ng, J.~Egedal, A.~Le, W.~Daughton, and L.~J. Chen, ``Kinetic structure of
  the electron diffusion region in antiparallel magnetic reconnection,'' {\em
  Phys. Rev. Lett.}, vol.~106, FEB 10 2011.

\bibitem{gurnett:1997}
D.~GURNETT and L.~FRANK, ``Region of intense plasma-wave turbulence on auroral
  field lines,'' {\em J. Geophys. Res.}, vol.~{82}, no.~{7}, pp.~{1031--1050},
  {1977}.

\bibitem{matsumoto:1994}
H.~MATSUMOTO, H.~KOJIMA, T.~MIYATAKE, Y.~OMURA, M.~OKADA, I.~NAGANO, and
  M.~TSUTSUI, ``Electrostatic solitary waves (esw) in the magnetotail - ben
  wave-forms observed by geotail,'' {\em Geophy. Res. Lett.}, vol.~{21},
  pp.~{2915--2918}, {DEC 15} {1994}.

\bibitem{ergun:2009}
R.~E. Ergun, L.~Andersson, J.~Tao, V.~Angelopoulos, J.~Bonnell, J.~P. McFadden,
  D.~E. Larson, S.~Eriksson, T.~Johansson, C.~M. Cully, D.~N. Newman, M.~V.
  Goldman, A.~Roux, O.~LeContel, K.~H. Glassmeier, and W.~Baumjohann,
  ``{Observations of Double Layers in Earth's Plasma Sheet},'' {\em Phys. Rev.
  Lett.}, vol.~{102}, {APR 17} {2009}.

\bibitem{andersson:2009}
L.~Andersson, R.~E. Ergun, J.~Tao, A.~Roux, O.~LeContel, V.~Angelopoulos,
  J.~Bonnell, J.~P. McFadden, D.~E. Larson, S.~Eriksson, T.~Johansson, C.~M.
  Cully, D.~L. Newman, M.~V. Goldman, K.~H. Glassmeier, and W.~Baumjohann,
  ``{New Features of Electron Phase Space Holes Observed by the THEMIS Mission
  (vol 102, art no 225004, 2009)},'' {\em Phys. Rev. Lett.}, vol.~{103}, {JUL
  31} {2009}.

\bibitem{mozer:2013}
F.~S. Mozer, S.~D. Bale, J.~W. Bonnell, C.~C. Chaston, I.~Roth, and J.~Wygant,
  ``{Megavolt Parallel Potentials Arising from Double-Layer Streams in the
  Earth's Outer Radiation Belt},'' {\em Phys. Rev. Lett.}, vol.~{111}, {DEC 2}
  {2013}.

\bibitem{li:2013}
T.~C. Li, J.~F. Drake, and M.~Swisdak, ``Coronal electron confinement by double
  layers,'' {\em ASTROPHYSICAL JOURNAL}, vol.~{778}, {DEC 1} {2013}.

\bibitem{li:2014}
T.~C. Li, J.~F. Drake, and M.~Swisdak, ``Dynamics of double layers, ion
  acceleration, and heat flux suppression during solar flares,'' {\em
  ASTROPHYSICAL JOURNAL}, vol.~{793}, {SEP 20} {2014}.

\bibitem{shay:2004}
M.~Shay, J.~Drake, M.~Swisdak, and B.~Rogers, ``{The scaling of embedded
  collisionless reconnection},'' {\em Phys. Plasmas}, vol.~{11},
  pp.~{2199--2213}, {MAY} {2004}.

\bibitem{shay:2001}
M.~Shay, J.~Drake, B.~Rogers, and R.~E. Denton, ``Alfvenic collisionless
  magnetic reconnection and the hall term,'' {\em J. Geophys. Res.}, vol.~106,
  pp.~3759--3772, MAR 1 2001.

\bibitem{le:2014}
A.~Le, J.~Egedal, J.~Ng, H.~Karimabadi, J.~Scudder, V.~Roytershteyn,
  W.~Daughton, and Y.~H. Liu, ``{Current sheets and pressure anisotropy in the
  reconnection exhaust},'' {\em Phys. Plasmas}, vol.~{21}, {JAN} {2014}.

\bibitem{buchner:1989}
J.~Buchner and L.~Zelenyi, ``Regular and chaotic charged-particle motion in
  magnetotail-like field reversals .1. basic theory of trapped motion,'' {\em
  J. Geophys. Res.}, vol.~94, pp.~11821--11842, SEP 1 1989.

\bibitem{dahlin:2014}
J.~T. Dahlin, J.~F. Drake, and M.~Swisdak, ``{The mechanisms of electron
  heating and acceleration during magnetic reconnection},'' {\em Phys.
  Plasmas}, vol.~{21}, {SEP} {2014}.

\bibitem{beresnyak:2014}
A.~Beresnyak and H.~Li, ``First order particle acceleration in
  magnetically-driven flows,'' {\em submitted to {\sl PRL},}, {2014}.

\bibitem{nagai:2001}
T.~Nagai, I.~Shinohara, M.~Fujimoto, M.~Hoshino, Y.~Saito, S.~Machida, and
  T.~Mukai, ``Geotail observations of the hall current system: Evidence of
  magnetic reconnection in the magnetotail,'' {\em J. Geophys. Res.}, vol.~106,
  pp.~25929--25949, NOV 1 2001.

\bibitem{asano:2008}
Y.~Asano, R.~Nakamura, I.~Shinohara, M.~Fujimoto, T.~Takada, W.~Baumjohann,
  C.~J. Owen, A.~N. Fazakerley, A.~Runov, T.~Nagai, E.~A. Lucek, and H.~Reme,
  ``{Electron flat-top distributions around the magnetic reconnection
  region},'' {\em J. Geophys. Res.}, vol.~{113}, {JAN 24} {2008}.

\bibitem{lin:1995}
L.~R. P., A.~K. A., A.~S., C.~C., C.~D., E.~R., L.~D., M.~J., M.~M., P.~G. K.,
  R.~H., B.~J. M., C.~J., C.~F., D.~C., W.~K. P., S.~T. R., H.~J., R.~J. C.,
  and P.~G., ``A 3-dimensional plasma and energetic particle investigation for
  the wind spacecraft,'' {\em Space Science Reviews}, vol.~71, pp.~125--153,
  FEB 1995.

\bibitem{oieroset:2002}
M.~{\O}ieroset, R.~Lin, and T.~Phan, ``Evidence for electron acceleration up to
  300 kev in the magnetic reconnection diffusion region of earth's
  magnetotail,'' {\em Phys. Rev. Lett.}, vol.~89, p.~195001, NOV 4 2002.

\bibitem{balogh:2001}
A.~{Balogh}, C.~M. {Carr}, M.~H. {Acu{\~n}a}, M.~W. {Dunlop}, T.~J. {Beek},
  P.~{Brown}, K.-H. {Forna{\c c}on}, E.~{Georgescu}, K.-H. {Glassmeier},
  J.~{Harris}, G.~{Musmann}, T.~{Oddy}, and K.~{Schwingenschuh}, ``{The Cluster
  Magnetic Field Investigation: overview of in-flight performance and initial
  results},'' {\em Annales Geophysicae}, vol.~19, pp.~1207--1217, 2001.

\bibitem{reme:2001}
H.~Reme, C.~Aoustin, M.~Bosqued, I.~Dandouras, B.~Lavraud, J.~Sauvaud,
  A.~Barthe, J.~Bouyssou, T.~Camus, O.~Coeur-Joly, A.~Cros, J.~Cuvilo,
  F.~Ducay, Y.~Garbarowitz, J.~Medale, E.~Penou, H.~Perrier, D.~Romefort,
  J.~Rouzaud, C.~Vallat, D.~Alcayde, C.~Jacquey, C.~Mazelle, C.~d'Uston,
  E.~Mobius, L.~Kistler, K.~Crocker, M.~Granoff, C.~Mouikis, M.~Popecki,
  M.~Vosbury, B.~Klecker, D.~Hovestadt, H.~Kucharek, E.~Kuenneth, G.~Paschmann,
  M.~Scholer, N.~Sckopke, E.~Seidenschwang, C.~Carlson, D.~Curtis, C.~Ingraham,
  R.~Lin, J.~McFadden, G.~Parks, T.~Phan, V.~Formisano, E.~Amata,
  M.~Bavassano-Cattaneo, P.~Baldetti, R.~Bruno, G.~Chionchio, A.~Di~Lellis,
  M.~Marcucci, G.~Pallocchia, A.~Korth, P.~Daly, B.~Graeve, H.~Rosenbauer,
  V.~Vasyliunas, M.~McCarthy, M.~Wilber, L.~Eliasson, R.~Lundin, S.~Olsen,
  E.~Shelley, S.~Fuselier, A.~Ghielmetti, W.~Lennartsson, C.~Escoubet,
  H.~Balsiger, R.~Friedel, J.~Cao, R.~Kovrazhkin, I.~Papamastorakis, R.~Pellat,
  J.~Scudder, and B.~Sonnerup, ``{First multispacecraft ion measurements in and
  near the Earth's magnetosphere with the identical Cluster ion spectrometry
  (CIS) experiment},'' {\em ANNALES GEOPHYSICAE}, vol.~{19}, pp.~{1303--1354},
  {OCT-DEC} {2001}.

\bibitem{johnstone:1997}
A.~D. Johnstone, C.~Alsop, S.~Burge, P.~J. Carter, A.~J. Coates, A.~J. Coker,
  A.~N. Fazakerley, M.~Grande, R.~A. Gowen, C.~Gurgiolo, B.~K. Hancock,
  B.~Narheim, A.~Preece, P.~H. Sheather, J.~D. Winningham, and R.~D. Woodliffe,
  ``{Peace: A plasma electron and current experiment},'' {\em {Space
  Sci.~Rev.}}, vol.~{79}, pp.~{351--398}, {JAN} {1997}.

\bibitem{borg:2012a}
A.~L. Borg, M.~G. G.~T. Taylor, and J.~P. Eastwood, ``{Observations of magnetic
  flux ropes during magnetic reconnection in the Earth's magnetotail},'' {\em
  ANNALES GEOPHYSICAE}, vol.~{30}, no.~{5}, pp.~{761--773}, {2012}.

\bibitem{egedal:2010grl}
J.~Egedal, A.~Le, Y.~Zhu, W.~Daughton, M.~{\O}ieroset, T.~Phan, R.~P. Lin, and
  J.~P. Eastwood, ``Cause of super-thermal electron heating during magnetotail
  reconnection,'' {\em Geophy. Res. Lett.}, vol.~37, MAY 28 2010.

\bibitem{wygant:2005}
J.~R. Wygant, C.~A. Cattell, R.~Lysak, Y.~Song, J.~Dombeck, J.~McFadden, F.~S.
  Mozer, C.~W. Carlson, G.~Parks, E.~A. Lucek, A.~Balogh, M.~Andre, H.~Reme,
  M.~Hesse, and C.~Mouikis, ``{Cluster observations of an intense normal
  component of the electric field at a thin reconnecting current sheet in the
  tail and its role in the shock-like acceleration of the ion fluid into the
  separatrix region},'' {\em J. Geophys. Res.}, vol.~{110}, {SEP 3} {2005}.

\bibitem{masuda:1994}
S.~Masuda, T.~Kosugi, H.~Hara, and Y.~Ogawaray, ``A loop top hard x-reay source
  i a compact solar-flare as evidence for magnetic reconnection,'' {\em
  Nature}, vol.~371, pp.~495--497, OCT 6 1994.

\bibitem{lin:2003}
R.~P. Lin, S.~Krucker, G.~J. Hurford, D.~M. Smidth, and H.~S. Hudson, ``Rhessi
  observations of particle acceleration and energy release in an intense solar
  gamma-ray line flare,'' {\em Astrophys. J.}, vol.~595, pp.~L69--L76, 2003.

\bibitem{drake:2005}
J.~F. Drake, M.~A. Shay, W.~Thongthai, and M.~Swisdak, ``Production of
  energetic electrons during magnetic reconnection,'' {\em Phys. Rev. Lett.},
  vol.~94, p.~095001, MAR 11 2005.

\bibitem{guo:2014}
{Fan Guo}, {Hui Li}, W.~Daughton, and {Yi-Hsin Liu}, ``{Formation of Hard Power
  Laws in the Energetic Particle Spectra Resulting from Relativistic Magnetic
  Reconnection},'' {\em Phys. Rev. Lett.}, vol.~{113}, p.~{155005 (5 pp.)}, {10
  Oct.} {2014}.

\bibitem{birn:2013}
J.~Birn, M.~Hesse, R.~Nakamura, and S.~Zaharia, ``{Particle acceleration in
  dipolarization events},'' {\em J. Geophys. Res.}, vol.~118, pp.~1960--1971,
  MAY 2013.

\bibitem{runov:2011}
A.~Runov, V.~Angelopoulos, X.-Z. Zhou, X.~J. Zhang, S.~Li, F.~Plaschke, and
  J.~Bonnell, ``{A THEMIS multicase study of dipolarization fronts in the
  magnetotail plasma sheet},'' {\em J. Geophys. Res.}, vol.~116, MAY 24 2011.

\bibitem{hazeltine:1992}
R.~D. Hazeltine and J.~D. Meiss, {\em Plasma Confinement}.
\newblock Addison-Wesley, 1992.

\end{thebibliography}

\end{document}